\documentclass[10pt,preprint]{aastex}
\usepackage{natbib}
\usepackage{graphicx}
\usepackage{epsf}

\bibpunct{(}{)}{;}{a}{}{,} 

\shorttitle{Close-up Extinction Maps of the Pipe Nebula}
\shortauthors{Rom\'an-Z\'u\~niga et al.}

\begin{document}


\title{Deep Near-Infrared Survey of Dense Cores in the Pipe Nebula II:
Data, Methods, and Dust Extinction Maps}


\author{Carlos G. Rom\'an-Z\'u\~niga\altaffilmark{1}, Jo\~ao F. Alves\altaffilmark{2}, Charles J. Lada\altaffilmark{3} and
Marco Lombardi\altaffilmark{4}}
\altaffiltext{1}{Centro Astron\'omico Hispano Alem\'an \/ Instituto de Astrof\'isica de
Andaluc\'ia (IAA-CSIC), Glorieta de la Astronom\'ia, S/N, Granada, Spain, 18008}
\altaffiltext{2}{Institute of Astronomy, University of Vienna, T\"urkenschanzstr. 17, 1180 Vienna, Austria}
\altaffiltext{3}{Harvard Smithsonian Center for Astrophysics, 60 Garden Street, Cambridge MA 02138}
\altaffiltext{4}{European Southern Observatory, Karl-Schwarzschild-Strasse 2, Garching 85748, Germany}



\begin{abstract}

We present a new set of high resolution dust extinction maps of the nearby and essentially starless Pipe Nebula molecular cloud. The maps were  constructed from a concerted deep near-infrared imaging survey with the ESO-VLT, ESO-NTT, CAHA 3.5m telescopes, and 2MASS data. The new maps have a resolution three times higher than the previous extinction map of this cloud by Lombardi et al. (2006) and are able to resolve structure down to 2600 AU. We detect 244 significant extinction peaks across the cloud. These peaks have masses between 0.1 and 18.4 M$_\odot$, diameters between 1.2 and 5.7$\times 10^4$ AU (0.06 and 0.28 pc), and mean densities of about 10$^4$ cm$^{-3}$, all in good agreement with previous results. From the analysis of the Mean Surface Density of Companions we find a well defined scale near $1.4\times10^4$ AU below which we detect a significant decrease in structure of the cloud. This scale is smaller than the Jeans Length calculated from the mean density of the peaks. The surface density of peaks is not uniform but instead it displays clustering. Extinction peaks in the Pipe Nebula appear to have a spatial distribution similar to the stars in Taurus, suggesting that the spatial distribution of stars evolves directly from the primordial spatial distribution of high density material. 

\end{abstract}


\section{Introduction \label{s:int}}

A primary motivation for the study of molecular clouds at early stages of evolution is to understand the initial conditions that precede the process of star formation. Those initial conditions are necessary to understand the process of evolution of the clouds towards collapse into stars, and to improve numerical and analytical experiments that should lead to a predictive model of star formation. However, there are not many examples of clouds near or at primordial stages. That is why the Pipe Nebula is both an important and interesting target. It is one of the youngest clouds observable, possibly magnetically dominated \citep{Franco:2010aa}, with less than a handful of its dense cores being confirmed as currently forming stars \citep{Onishi:1999aa,Brooke:2007aa,Forbrich:2009ab}. The cloud is also located at a close distance ($\mbox{d}=130^{+13}_{-20}$ pc; \citet{Lombardi:2006aa}, hereafter LAL06), offering a great level of detail to observers. 

As clouds evolve, their structure is modified. The most direct way to determine the (projected) structure of a molecular cloud is to determine the variation of an observable column density across the cloud. One method that has proven to be both reliable and accurate is to use dust extinction as a proxy for total mass (i.e. gas column density). One successful approach to this method is the near infrared color excess technique (\citealt{Lada:1994aa}), revised and optimized by \citeauthor{Lombardi:2001aa}(\citeyear{Lombardi:2001aa}; hereafter LA01). In this technique, stars located in the background of clouds provide individual pencil-beam measurements of reddening towards a given line of sight. The colors of stars are averaged, weighted, and convolved into a uniform map of dust column density using a canonical reddening law and a statistical optimization method. In their study of the Pipe Nebula, LAL06 applied the color excess method to a sample of almost 4.5 million sources in the region catalogued by the Two-Mass All-Sky survey (2MASS). This allowed them to construct a detailed map of the Pipe Nebula with a spatial resolution of 1 arcminute, and an overall detection level of dust extinction of $A_V=0.5$ mag. The analysis of the map revealed a large number of column density peaks, resulting in a list of 134 candidate dense cores. The Pipe cores may also be the direct predecessors of a stellar population with an initial mass function similar to that of the Trapezium cluster (\citealt{Alves:2007aa}; \citealt{Rathborne:2009aa}, hereafter RLA09). However, while the map of LAL06 permitted a great ``in bulk" view of the core population, higher spatial resolution is needed to resolve the internal structure of the cores themselves. The analysis of the internal structure of individual cores may help to determine their properties and their possible stage of evolution towards collapse into stars (e.g. \citealt{Lada:2004aa}). To date, such analysis has been done for only two cores in the Pipe Nebula: Barnard 68 \citep{Alves:2001aa} and FeSt-1457 (\citealt{Kandori:2005aa}). 

This paper describes a new deep, near-infrared imaging survey of the Pipe Nebula. These new observations are 4 to 5 magnitudes deeper than 2MASS, resulting in a significant increase of the density of background sources per unit area, allowing to construct extinction maps with higher spatial resolution and to resolve the structure of the cloud in greater detail. The observations for this survey include data from three ground based near-infrared facilities: the Infrared Spectrometer and Array Camera (ISAAC) at the Very Large Telescope (VLT), the Son of Isaac (SOFI) camera at the New Technology Telescope (NTT) --both part of the European Southern Observatory (ESO)-- and the OMEGA 2000 wide field camera at the 3.5m telescope at the Centro Astron\'omico Hispano Alem\'an in Calar Alto. In this paper we describe the observations and present the data in the form of extinction maps for individual and combined fields. This paper is also a follow up to a previous study of the structure of the Barnard 59 star forming region \citep[][hereafter Paper I]{Roman-Zuniga:2009aa}, made with a subset of this survey. In this paper we pay special attention to the nomenclature of the denser regions in the cloud: we take the conservative approach of using ``extinction peaks" to define these dense regions, as extinction peaks are the direct observables from the extinction map. We do not call them ``cores" because we have not measured radial velocity for every peak, and therefore we cannot merge  peaks into ``cores", as it was done in the analysis of RLA09. 

This paper is organized as follows: in section \ref{s:obs} we make a general description of the observations, and describe our data reduction process. In section \ref{s:maps} we present the extinction maps and describe their construction. In section \ref{s:analysis} we make a brief analysis of the structure in the maps and the detection of column density peaks. In section \ref{s:results} we present our results and compare them with the studies of the 2MASS map of LAL06. Finally, a discussion and a summary of the main results are presented in sections \ref{s:discussion} and \ref{s:summary}, respectively.

\section{Observations and Data Reduction \label{s:obs}}

A list of all fields observed and considered for final analysis can be consulted in the Appendix ($\S$\ref{s:app:observations}, Table \ref{tab:obs}). The table lists the field identification, the center of field positions, observation date, filter, an estimate of the seeing based on the average FWHM of the stars detected in each field, and the peak values for the brightness distributions, which are a good measurement of the sensitivity limits achieved. The location and coverage of each field is indicated in Figure \ref{fig:FOV_MAP}. In what follows we describe the observations and the data reduction process, including the construction of the photometric catalogs used to construct the dust extinction maps.

\subsection{ESO \label{s:obs:ss:eso}}

The main observations of the survey were made with the Infrared Spectrometer and Array Camera (ISAAC) and the Son of ISAAC (SOFI) near-infrared imagers, mounted, respectively, at the UT3 8.2m unit of the Very Large Telescope (VLT) array at Cerro Paranal and the 3.5m New Technology Telescope (NTT) at La Silla, both part of the European Southern Observatory (ESO) in Chile.  The two observing runs were completed with uniformly good weather in the summer months of 2001 and 2002.  Two fields, FeSt 1-457 and B68, were observed previously in 2000 under similar conditions and requirements. All the data is currently available at the ESO archives. The imager SOFI at the NTT 3.5m telescope, presents a field of view (FOV) of $5\arcmin\times 5\arcmin$ with plate scale of 0.288 $\mbox{arcsec pix}^{-1}$, almost seven times the angular resolution of 2MASS, and higher sensitivity: achieved limits are estimated to be about 4 to 5 magnitudes deeper than 2MASS in each band. Such characteristics are enough to resolve the densely crowded Galactic Bulge field in the background of the Pipe Nebula and to penetrate in regions with up to about 50 mag of visual extinction. The FOV of SOFI is ample enough to allow to completely contain some of the larger dense regions in the Pipe Nebula (R$\sim$0.1-0.15 pc), with the exception of the central core in Barnard 59, which nearly doubles the size of the FOV. A total of 55 fields in the cloud were successfully observed in $H$ and $K_s$, with a few also observed in $J$. In addition, a control field located approximately 1$^\circ$ west of Barnard 59 was observed with similar conditions in order to determine the intrinsic background field colors, essential to determine absolute dust extinction values. The ISAAC observations were intended to fully resolve the centers of the most dense and obscured regions. A total of seven fields were observed with ISAAC during the same observing seasons in $H$ and $K_s$ at a resolution of 0.144 $\mbox{arcsec pix}^{-1}$ with a FOV a quarter of the size of that of SOFI. Interestingly, the ISAAC fields observed in the center of Barnard 59 were capable of detecting only a handful of the individual sources in and behind the core. This shows that extinction values above $A_V=60$ mag are close to the limit of the penetrating power in the near-infrared even for an 8m class telescope. As mentioned before, space based mid-infrared observations were required to resolve more of the highly obscured background sources in B59 (\citealt{Brooke:2007aa}; \citealt{Roman-Zuniga:2007aa}; Paper I).

\subsection{CAHA \label{s:obs:ss:caha}}

The CAHA 3.5m observations were done with the OMEGA 2000 camera, which has a wide field of view of 15$\arcmin$. A total of 21 fields, complementary to the ESO survey, were observed during june 2007 and june 2008 with acceptable weather. The fields observed with the CAHA 3.5m telescope were selected to cover the surrounding areas in the densest region near the ``Pipe Molecular Ring" \citep{Muench:2007aa} in the central part of the cloud, and also to obtain fresh data in a number of fields at the westernmost regions of the cloud (see Figure \ref{fig:FOV_MAP}). The resolution of OMEGA 2000 at the 3.5m telescope is 0.45 $\mbox{arcsec pix}^{-1}$ and the sensitivity is expected to be equivalent to that of SOFI. However, even with excellent weather and instrumental conditions at Calar Alto, the low inclination of the Pipe Nebula at the latitude of the observatory resulted in large seeing values ($1\farcs6 \pm 0.2$ in average) and some internal reflection effects that affected the colors of sources near the edges of the detector. See also $\S$\ref{s:maps:ss:nicer}.

\subsection{Pipeline Reduction \label{s:obs:ss:red}}

Data from the three survey groups were reduced with modified versions of the FLAMINGOS near-infrared reduction 
and photometry/astrometry pipelines, which are built in the standard \texttt{IRAF} Command Language environment. 

One pipeline \citep[see][]{Roman-Zuniga:2006aa} processes all raw frames by subtracting darks and dividing by flat fields, improving signal to noise ratios by means of a two pass sky subtraction method, and combining reduced frames with an optimized centroid offset calculation.

For the ESO survey data, slight modifications were implemented to the reduction pipeline in order to account for some --well documented-- instrumental biases: For the SOFI-NTT data, we took into account corrections for instrumental crosstalk using the IRAF task provided in the ESO SOFI webpages.  Non-linearity corrections were applied by using the coefficients listed by \citet{Tinney:2003aa}. Finally, we used the dome flats and illumination correction field sets provided by ESO to avoid well known ``shade'' effects often encountered with sky and super-sky flat fields. In the case of the ISAAC-NTT fields, master darks, twilight flats and illumination field frames were created with the aid of the ISAAC pipelines provided in the ESO webpages, previous to the batch reduction with our pipelines. For the Calar Alto data, the reduction was performed using our pipeline in standard mode. We used dark frames and dome flats obtained within 48 hours of each observation. No other specific instrumental corrections were applied.

The final combined product images were then analyzed by a second pipeline, \citep[see][]{Levine:2006ab}, which identifies all possible sources from a given field using the \texttt{SExtractor} algorithm \citep{Bertin:1996aa}, applies PSF photometry, calibrates observed magnitudes to a constant zero point and finds an accurate astrometric solution. Both photometry and astrometry of SOFI and OMEGA 2000 data products were calibrated with respect to 2MASS catalogs retrieved from the All Sky Data Release databases. ISAAC calibrations were obtained by bootstrapping to the calibrated SOFI data. The final photometry catalogs, containing either $J$, $H$, and $K_s$, or $H$ and $K_s$ photometry were prepared by a routine that makes use of the routine \texttt{CCXYMATCH}. In the case of overlapping fields and fields with observations done in more than one instrument, the catalogs were put together with our own matching routines, designed to list preferentially a higher quality observation (e.g. ISAAC) over a lower quality one in the overlapping areas. We prepared joint catalogs to construct multiple-field extinction maps for a limited number of cases. In the cases of FeSt 1-457, and SOFI fields 29, 31, and 45, we used hybrid SOFI+ISAAC catalogs.

\section{High Resolution Dust Extinction Maps \label{s:maps}}

The ESO and CAHA observations provide a vast enhancement in photometric depth and sensitivity toward the fields observed. The number of sources detected in each field is up to ten times larger than 2MASS, which results in the ability to increase the spatial resolution used to map the dust column density: with a greater source density, one can reduce the diameter of the spatial filter and have equivalent or better statistics, with an improved signal to noise ratio. 

\subsection{Near Infrared Color Excess Method \label{s:maps:ss:nicer}}

The maps were constructed with the Near Infrared Color Excess Revised (NICER) technique \citep{Lombardi:2001aa}. NICER is an optimized multi-band technique to estimate extinction from the infrared excess calculated from observations at three different bands --in our case $J$, $H$, and $K_s$, which allow to use two independent colors. Choosing $J-H$ and $H-K_s$, the estimator of the extinction, $A_V$ is of the form:

\begin{equation}
A_V(s) = a+ b_1[E(J-H)]+b_2[E(H-K_s)]
\end{equation}

\noindent where the coefficients, $a$, $b_1$ and $b_2$ can be determined by supposing that $A_V$ is an unbiased estimator and that it has minimum variance. The first condition is expressed as $b1/C_1+b2/C_2=1$, where, in our case, the coefficients $C_1=9.35$ and $C_2=16.23$ are from the reddening law of \citet{Roman-Zuniga:2007aa}. The second condition is expressed as $a+b_1(J-H)_0 + b_2(H-K_s)_0=0$, where $(J-H)_0$ and $(H-K)_0$ are the intrinsic values of the colors, estimated from an off-cloud control field with negligible extinction (see Figure 1 and Table 1). The second condition is granted through the minimization of the variance of $A_V$, expressed in terms of the scatter of the intrinsic colors and the photometric errors (please see LA01 for details). We estimated the intrinsic values of background sources in the control field as the median value along a fiducial line represented by the average colors in bins of equal size in $K_s$, for $K_s<17.0$ mag. The intrinsic colors are practically constant across the control field area, with scatter comparable to or smaller than the typical color photometric uncertainty. This is expected, because the field population towards the Galactic Bulge is mostly composed of red giant stars, which have a very small intrinsic color dispersion in the near-infrared. However, the source catalogs also have to be restricted to reject any stars with intrinsic color excess, like in the case of OH IR stars, as discussed by LAL06. We selected infrared excess stars two ways: for sources having J, H, and K photometry we selected out those falling to the right of the line $H-K_s=1.692(J-H)$. For sources not having a J band observation we cannot guarantee that they will not have an intrinsic excess, specially if $H-K_s>1$. However those cases should be rare in background giant sources. For this reason, we limited our catalogs to stars with $K_s>10.0$; this restriction removes contamination from foreground stars and most contamination from OH-IR stars (see LAL06).

To make a map, the extinction measurements for individual sources have to be spatially smoothed. In our case, we used a Gaussian filter and Nyquist sampling. At each position (line of sight) in the map, each star falling within the beam is given a weight calculated from a Gaussian function, and the inverse of the variance squared. Then, the value of $A_V$ at the map position is calculated as the weighted median of all possible values (please see LA01, sect 3.2 for details). The uncertainty in the measurement, $\sigma_{A_V}$, which determines the noise per pixel, was calculated as $\sqrt{\sigma^2(A_V)/N}$.

We do not have ESO $J$ band observations available for a majority of the ESO fields. Therefore, for many ESO sources, extinction was determined from only one color, $H-K_s$, just as in the original color excess technique (NICE, \citealt{Lada:1994aa}). The difference is that using the NICER method we are able to use the error associated with a measurement to keep track of its corresponding weight. For example, by setting the photometric uncertainty of stars with no $J$ measurement at a very large value (99.999) in that band, NICER automatically discards that $J-H$ measurement and either accepts or rejects a value of $A_V$ for a star given the weight associated with the value of its $H-K_s$ color alone.

The maps we present in this paper are constructed with a Gaussian filter with a FWHM of 20$\arcsec$ and Nyquist sampling (i.e. pixels on the map have a width of $10\arcsec$). This represents a resolution three times higher than the one achieved in the large scale map of LAL06. This is approximately equal to $1/10$ of the median Jeans length across the cloud (0.2 pc; RLA09). 

In the case of the CAHA observations we had three bands available, $H$, $J$, and $K_s$, thus two colors $c_1=H-K_s$, $c_2=J-K_s$, and a pixel sampling not too different from SOFI. However, the Pipe Nebula is a target that does not reach a very high elevation at Calar Alto, and the inclination of the telescope was large enough to cause internal reflections in the camera. These reflections affected the colors of stars near the corners of the frames. We decided to mask out the frame corners by using sources within a circular area with a radius smaller to the diagonal length of the field. However, because each OMEGA 2000 field contained an average of 3.5$\times 10^4$ detected sources, we were still able to obtain uniformly covered maps with a very acceptable number of 15 to 60 sources per pixel and an average completeness limit of $K_s=16.75$ after applying the restrictions, which was enough to resolve structures in all regions of the Pipe (with the exception of B59) at the nominal resolution of 20$\arcsec$.

\subsection{Maps for Individual Fields\label{s:maps:s:atlas}}

We constructed individual dust extinction maps for each of the SOFI and CAHA fields observed. In the case of the few deep ISAAC fields observed, we merged the catalogs with the corresponding ones from SOFI before proceeding to the map construction following the procedure described below. The individual ESO and CAHA maps constitute an high resolution dust extinction map atlas of the Pipe Nebula which we present as a separate document, accessible in the electronic (online) version of the paper. The dynamical range of the new maps covers column densities between $10^{21}\mbox{ and } 10^{23}\mbox{ cm}^{-2}$.

\subsection{Large Scale Maps \label{s:maps:s:largemaps}}

We merged overlapping ISAAC, SOFI and CAHA catalogs to form larger pieces which were subsequently merged onto five large ``bed" catalogs that defined the main regions of the cloud. This way, we constructed five large scale maps comprising the main regions of the Pipe --namely ``Barnard 59", ``The Stem", ``The Shank", ``The Bowl" and ``The Smoke". A diagram showing the extent of each region is presented in Figure \ref{fig:MAP_REGS}. The five large scale extinction maps are shown in figures \ref{fig:LMAP_B59} to \ref{fig:LMAP_SMOKE}. In Table \ref{tab:mapareas} we list the limits defining the five areas, the fields included in each one of the maps and the number of sources included from each of the surveys. The hybrid catalogs were constructed by joining the photometry lists from the various surveys in a progressive sequence: ISAAC catalogs were merged with the corresponding SOFI catalogs by using a matching algorithm that preferentially listed ISAAC over SOFI detections. For those matches with a SOFI $J$ band detection, we used that value to replace the the null $J$ band value from ISAAC (e.g. FeSt 1-457 field); this resulted in a small but useful reduction of the scatter as a function of extinction ($\sigma_{A_V}$ vs $A_V$). Positional matching between NTT and VLT observations was calculated within a tolerance radius of $0\farcs24$, i.e. twice the average positional uncertainty of the NTT astrometric solutions. Then, when possible, we merged coincident ESO and CAHA fields using a slightly larger tolerance radius of $0\farcs3$ (to compensate for the slightly lower resolution of the OMEGA 2000 respect to SOFI). A final ``high resolution" catalog was constructed by putting together all ESO-CAHA merges, and merging it to a 2MASS ``bed" catalog obtained from the All Sky Data Release. The 2MASS ``bed" catalogs define the extent of each map, and excluded sources with photometric flags `U' and `X' which indicate upper limits and defective observations, respectively. Positional matching between the ESO-CAHA catalogs and the 2MASS bed catalogs was determined using a tolerance radius of $0\farcs36$ (three times the average uncertainty of the ESO-NTT astrometric solutions), and we rejected the 2MASS matching sources except in the cases where an ESO or CAHA source was saturated. The infrared excess restriction applied to the 2MASS sources is the same as the one described in $\S$\ref{s:maps:ss:nicer}. 

We determined that a small color correction had to be applied to correct for the differences between the SOFI/ISAAC, OMEGA-2000 and 2MASS passbands. This correction was determined by plotting the differences $\Delta(H-K_s)=(H-K_s)-(H-K_s)_{2MASS}$ for ESO or CAHA versus $(J-H)_{2MASS}$ and applying a linear least squares fit. The slopes did not vary much between ESO and CAHA fields. We estimated an average correction of $\Delta(H-K_s)=-0.08+0.07(J-H)_{2MASS}$ and applied it directly to the individual $H-K_s$ colors of sources entering the NICER calculations. 

\section{Analysis of Large Scale Maps \label{s:analysis}}

One of the goals of this study is to determine if the increase in resolution has any direct implications in the determination of the structure of the cloud and the properties of identified dense regions. We also need to determine if the increase in spatial resolution has any significant impact on our ability to identify high column density regions. To tackle such aspects, we analyzed our maps following a technique equivalent to that followed by LAL06 and RAL09 to analyze the 2MASS map and then compared our results to theirs. 

\subsection{Identification of Extinction Peaks \label{s:analysis:ss:clumps}}

The maps of figures \ref{fig:LMAP_B59} to \ref{fig:LMAP_SMOKE} show that there is abundant low density material between $A_V=2\pm1$ and $A_V=6\pm1$ mag with a rough filamentary morphology. This is probably tracing a more diffuse medium, as most significant extinction peaks appear to be embedded in these relatively large, low density structures. The correct detection and delimitation of individual features is complicated by this aspect, because we need to be careful at defining clearly the boundary of a feature projected against an extended emission structure. In some cases two or more peaks are located close to each other and are projected toward the same filament or wisp of the cloud, making it difficult to determine their individual boundaries. Because we estimate sizes and masses of features from those boundaries, crowding and overlapping may ultimately play a crucial role in the correct determination of a dense core mass function for the cloud, although in the Pipe Nebula, crowding is a relatively small effect \citep{Kainulainen:2009aa} due to the proximity and particular layout the cloud. In the previous studies of extinction maps of the Pipe Nebula, the detection of individual extinction peaks was optimized in two steps: 1) filtering the local background structures and 2) extracting individual features by means of a two-dimensional peak finding algorithm. We also follow these steps in our analysis, as described below:

\subsubsection{Filtering and Detection \label{s:analysis:ss:clumps:sss:detection}}

The Multi-scale Vision Model algorithm \citep{Rue:1997aa,Bijaoui:1997aa} translated into a computer routine by B. Vandame (personal communication) works well for our extinction maps because it is optimized to extract extinction peaks projected against an extended background, a picture that fits well with a collection of dense gas cores embedded in a large cloud. The algorithm assumes that significant extinction peaks are compact, coherent structures that will remain significant after moderate changes in resolution. Large scale structures such as filaments and ``smooth hills", are effectively filtered out by the code as they usually are too extended.

In our extinction maps we applied the wavelet filtering by isolating features at four scales, of sizes $2^{l}\cdot s_p$, where $s_p$ is the spatial resolution scale of $20\arcsec$ and $l=\{1,..,4\}$, i.e. $0\farcm 7$, $1\farcm 3$, $2\farcm 7$, and $5\farcm 3$. At each of these scales, the program detects local maxima among pixels with values higher than a threshold (3-sigma, in our case); in the wavelet transform space, pixels associated with local maxima are organized in domains; the program then constructs a series of trees of inter-scale connectivity, and reconstructs the image by summing across all the scales with the help of a regularization (minimum energy loss) formulation. We checked that the filtering at these four scales was sufficient by subtracting the filtered images from the originals. In all cases the residual images only presented noise and some low level emission features, coincident with faint, wispy structures visible in the large scale maps (for an example, see paper I, Fig. 5). 

We identified individual extinction peaks by running the \texttt{CLUMPFIND-2D} (hereafter\ CLF2D) algorithm \citep{Williams:1994aa} on the wavelet filtered maps. The algorithm detects individual peaks as local maxima and encloses adjacent regions associated with them, defining individual boundaries or ``clumps"\footnote{The word ``clump" is used in a generic way by Williams et al. to describe their algorithm, but it should not be confounded with other definitions in the literature (especially in molecular emission map analyses), in which ``clumps" are considered to be relatively large structures containing groups of cores. We avoid the use of the word ``clump" in the present study.}. The program makes use of contour levels based on user defined intervals of constant flux. In our case, the noise amplitude, $\sigma_{A_V}$, as a function of extinction was found to be relatively uniform in the areas of low extinction (mostly covered by 2MASS data) in the raw maps. In the Stem, Shank, Bowl and Smoke regions, the noise level has an average value of 0.33 mag between $0.0<A_V<30.0$ mag. The noise level is lower in regions covered by SOFI or CAHA observations), then it rises smoothly to an average of 0.55 mag as $A_V$ increases to peak values of 40 to 50 mag per pixel (Bowl region). Following this behavior, we constructed the contour level set with intervals starting at $A_V=0.99$ mag contour, defined as a ``base" level equal to 3 times the average noise per pixel in low extinction ($A_V<5$ mag) regions, followed by 5$\sigma_{A_V}$ intervals (1.65 mag) within $1<A_V<30$ mag (low and moderate extinction), and ending with 2.0 mag intervals for $A_V>30$. The extra $1\sigma$ interval in the last segment is used to compensate for the increase of the noise amplitude at high extinction regimes. This way CLF2D finds all extinction peaks in the wavelet filtered maps and works its way down the contour levels, separating the peaks by determining which pixels most likely belong to each of them (by means of a ``friends of friends" algorithm) and finally determining the peak boundaries at the 3$\sigma$ ``base" level. We restricted the detection to only include peaks enclosing a  minimum of 30 contiguous pixels with values above the ``base" level; this is equivalent to saying that we only accepted as significant those features with equivalent diameters larger than 60$\arcsec$ (see below), i.e. about 1/8 of the average Jeans length in the cloud, estimated to be $\sim 0.2$ pc or $5\farcm 3$. Our choice of parameters should be considered to be on the conservative side, but we have to take into account that the map sensitivity is lower in the regions outside the ESO and CAHA fields, and a higher threshold helps to compensate partially for this effect. 

\subsubsection{Properties of Extinction Peaks \label{s:analysis:ss:clumps:sss:properties}}

For each of the extinction peaks identified, CLF2D defines an optimized boundary. The size of a feature is defined as the equivalent radius of the area within the boundary. Mass was estimated by summing the background corrected total extinction in pixels within the boundary. The conversion to mass, assuming a standard value for the gas to dust ratio $N_H/A_V = 2.0\times10^{21}\mathrm{cm}^{-2}$, is given as:

\begin{equation}
\frac{M_{peak}}{\mathrm{\ [M]_\odot}} = 1.28\times 10^{-10}\left (\frac{\theta}{\arcsec}\right )^2 \left(\frac{D_{cloud}}{\mathrm{pc}}\right)^2 \sum_{i=1}^{N}{(A_V)_i}\ M_\odot,
\end{equation}

\noindent where $\sum_{i=1}^{N}{(A_V)_i}$ adds the contribution of $N$ pixels within the feature boundary in the wavelet filtered map, $\theta$ is the beam size, and $D_{cloud}$ is the distance to the cloud. In our case, $D=130$ pc and $\theta =20\arcsec$. Using the values for mass and equivalent radii, we made estimates of the average density of the peaks, $\bar{n}=3M/4\pi\mu m_HR^3$. 
Uncertainties were calculated by adding random noise to the each of the five wavelet filtered maps (this was done following the $\sigma_{A_V}\mbox{ vs. }A_V$ behavior described in $\S$\ref{s:analysis:ss:clumps:sss:detection}), and then running CLF2D, to register the differences in mass and radius for all peaks that had a matched detection with the original map. We repeated this process 25 times, and then we calculated the mean deviation from the original values. The resultant uncertainties are 11.7\%, 12.5\% and 23.2\% in mass, radius, and density, respectively. 

Assuming a gas temperature of 10 K (consistent with estimates of $T_K$ from pointed observations of ammonia emission by RLA09) we calculated the local Jeans length, $L_J$ for each feature as:

\begin{equation}\label{eqn:LJL}
\frac{L_J}{\mathrm{\ [pc]}} = 0.2\left (\frac{T}{10\mathrm{K}}\right )^{1/2} \left ( \frac{\bar{n}}{10^4\mathrm{\ cm}^{-3}}\right )^{-1/2}. 
\end{equation}

\subsubsection{Resolving higher column density structures \label{s:analysis:ss:clumps:sss:usm}}

The depth and resolution of the 2MASS survey allowed LAL06 to produce a map with reliable column density measurements up to a limit of $A_V\approx 25$ mag. One of the main goals of our high resolution survey is to improve on this limit. Our high resolution maps now resolve the densest parts of the cloud and increase the previous peak values by a factor of up to 4. It is important to assess how significant are these changes for the physical characterization of the prestellar structure. 

In order to assess this effect, we applied a simple unsharp masking test to the large scale maps. This technique is frequently used in astronomical imaging to enhance or to determine the significance of detected features (e.g. \cite{Fabian:2006aa}, \cite{Moriarty-Schieven:2006aa}). For our maps, the masking is done in the following manner: first we convolved the maps with a , 60$\arcsec$ Gaussian beam, i.e. 3 times larger than the nominal resolution. These smoothed images are taken as being local equivalents to the map of LAL06. Then we subtracted these smoothed from the original images, resulting in frames with values fluctuating mostly around zero (with both positive and negative values), except on the centers of the denser peaks, where the residuals are relatively large and positive. We determined the mass of the residual and of the core in the 60$\arcsec$ convolved maps using the formulation described in $\S$\ref{s:analysis:ss:clumps:sss:properties}. The results are summarized in Figure \ref{fig:unsharp}, where we plot the ratio of the masses for positionally coincident features in the residual and the low resolution image. The ratios of the masses show clearly how the residual pixels only account, in most cases for a 5 to 30 percent of the total mass of the feature, with the larger differences being for the less massive peaks. Thus, it is unlikely that masses of peaks in the 2MASS map of \citeauthor{Alves:2007aa} were underestimated significantly because only the centers of the densest peaks were not resolved completely. Likewise, we should not expect to find a significant number of new high extinction peaks  on the high resolution maps, as the low extinction parts of these peaks should have been easily detected in the 2MASS map.

\section{Results \label{s:results}}

A total of 244 significant extinction peaks were found with our parameter restrictions within the areas covered by the five large scale maps. All these peaks are significant in size and column density with respect to the local background. However, we removed from the detection list seven peaks with marginal values of size and maximum extinction because they coincide in position with extinction peaks in the 2MASS map for which C$^{18}$O(1-0) emission could not be detected. This leaves a total of 237 peaks for analysis. There are 36 peaks that do not have a matching position in the RLA09 list, and they are identified as new detections. In Table \ref{tab:peaks} we list the position, size and mass for 220 peaks detected in the Stem, Shank, Bowl and Smoke maps. The 17 remaining peaks are already listed in Table 2 of Paper I.

\subsection{Global Properties \label{s:results:ss:peakproperties}}

A total of 90 individual extinction peaks in our list would fall below RLA09 detection threshold. This is equivalent to saying that about 37\% of the substructure we found in the new maps is below the minimum area threshold used by RLA09 in their analysis of the 2MASS map. This represents a significant increase in the detection of small scale structures. We compared the diameters and separations for our list of extinction peaks. The comparison is summarized in Figure \ref{fig:pixelcomp}, where we used astronomical units as well as pixel width (10$\arcsec$) as units of measurement. The peak to peak separations and peak diameters have median values equivalent to $1.6\mbox{ and } 2.3\times 10^4$ AU ($0.08$ pc and $0.11$ pc), respectively, which are near half of the median Jeans length at $T=10$ K ($3.7\times 10^4$ AU or $0.18\pm0.05$ pc). Notice that the distribution of peak separations decreases sharply below the median value, indicating that we do not detect a significant number of extinction peaks below that limit. We come to this point later in $\S$\ref{s:results:ss:peakspatialdist} and $\S$\ref{s:discussion}. The distribution of peak diameters shows that a majority of the extinction peaks appear to have sub-Jeans sizes, but it also shows that in most cases they are significant with respect to the minimum detection diameter of $8.3\times 10^3$ AU (0.04 pc). This confirms that at the resolution of our maps, we are measuring structures that have dimensions below the typical scale of thermal fragmentation for a significant number of cases. Moreover, in Figure \ref{fig:pixelcomp} we also show, for comparison, the detection threshold for the analysis of the 2MASS map, showing that the high resolution of our maps resulted in an effective increase in the detection of small scale structure, which was not resolved before.

In Figure \ref{fig:peak_mass_distribution} we present the mass distribution for the extinction peaks. The light colored solid curve, is the probability density function constructed with a Gaussian kernel with a width of 0.2 in the $\log{\mathrm{mass}}$ scale. The broken, dark colored line, is the best fit for the Trapezium IMF of \citet{Muench:2002aa} scaled by a constant (horizontal) factor in mass. The shape of the mass distribution shows a clear departure from a single power law at high mass. This break is  above the mass completeness limit 0.2 M$_\odot$ (see Appendix \ref{s:app:completeness}). Provided that each extinction peak would end up forming one star (an unlikely scenario, see $\S$\ref{s:results:ss:mergedcores} and $\S$\ref{s:discussion}), the mass scaling factor with respect to the Trapezium IMF is $1.6\pm0.1$, which, following the reasoning of \citet{Alves:2007aa}, translates into a star forming efficiency of $0.63\pm0.04$. 

In the top panel of Figure \ref{fig:mrplotdens} we show the mass vs. radius relationship considering all extinction peaks detected. The least square fit to the data points show that we agree well with the power law exponent of 2.6 found by \citet{Lada:2008aa}. The bottom panel shows the distribution of mean densities; the mean value for the extinction peaks is 1.1$\times10^4$ cm$^{-3}$, in good agreement with \citet{Lada:2008aa}.

\subsection{Spatial Distribution \label{s:results:ss:peakspatialdist}}

We calculated the Mean Surface Density of Companions (MSDC), $\Sigma_\theta$, for extinction peaks. The MSDC have been used extensively to discuss the spatial distribution of young stars in a number of regions, particularly in Taurus \citep{Gomez:1993aa,Larson:1995sr,Simon:1997aa,Hartmann:2002bj,Kraus:2008aa}.  Following \cite{Simon:1997aa}, we determined, for a list of $N_{peaks}$ peaks, the projected distances or separations, $\theta$, from each to every other feature. Then we organized them in bins (annuli) of 0.2 dex in the range $-2.0<\log{\theta\ ^\circ}<2.0$. Then $\Sigma_\theta$ was calculated by dividing the number of separations in each bin, $N_p(\theta)$, by the area of each annulus, and then normalizing by the total number of features. Another way to calculate $\Sigma_\theta$ is to use the two-point correlation function (TPCF), $W(\theta)$ \citep{Peebles:1973aa,Hewett:1982aa}, which compares the measured $\Sigma_\theta$ with that for a random, uniform distribution of points in the plane. We calculated 

\begin{equation}
W(\theta)=\frac{N_p(\theta)}{N_r(\theta)} - 1,
\end{equation}

\noindent where $N_r(\theta)$ is the number of separations per bin for a collection of points distributed at random positions over the same area as the survey, $A_{map}$. In our case, $A_{map}$ was chosen as a rectangle of $8\times 6(^\circ)^2$ which encloses all the regions in our maps in the RA-Dec plane. We evaluated $N_r$ from the average of 5000 drawings made with a Monte-Carlo routine. This way

\begin{equation}
\Sigma_\theta(TPCF)=(\frac{N_{peaks}}{A_{map}})(1+W(\theta)), 
\end{equation}

\noindent which can be compared with the MSDC estimated from direct counting.  The results are shown in Figure \ref{fig:stt}. We see that $\Sigma_\theta$ is anti-correlated to $\theta$, and that both estimates (MSDC and TPFC, white triangles and black squares, respectively) are consistent with each other in the range $-1.8<\log{\theta}<0.4$ (0.04 to 5.7 pc). The data points show that $\Sigma_\theta$ has a maximum near $\log{\theta ^\circ}=-1.5\approx \log{\theta\mbox{ pc}=-1.1}$, and then it flattens out. This break is below the median value of the local Jeans length but above our minimum detectable size scale. This suggests that peaks are genuine substructures extending below the thermal fragmentation length. Moreover, this maximum is equivalent to $1.45\times10^4\mbox{ AU}$ or 0.07 pc and it is not far from the median of distribution of peak to peak separations (16500 AU or 0.08 pc) in figure \ref{fig:pixelcomp}. This indicates that we detect a significant decrease in structure of the cloud at $1.45\pm0.2\times10^4\mbox{ AU}$, setting an observed limit of fragmentation in the Pipe Nebula.

For large values of $\theta$ ($\log{\theta}=0.4$ and above), the values of $\Sigma_\theta$ derived from the MSDC and the TPCF loose consistency because the separations start to be comparable to the angular size of the map itself and we loose information. The large ``bump" near $\log{\theta}=0.5$ ($\sim 3^\circ$, about 6.8 pc) in the large scale end of the TPCF is likely a signal (somewhat diluted) from the cloud itself, which at the largest separation scales is comparable to the size of the map. While the largest values of $\theta$ for the measured peaks are distributed around the median length of the cloud, the random placed peaks used to calculate the TPCF are distributed within the whole $8\times 6(^\circ)^2$ rectangle. We made a linear fit to the data points in the range $-1.5<\log{\theta}<0.4$. The resultant slope is -0.91$\pm$0.04, with a correlation (Pearson) coefficient close to 1.0. This near linear fit is indicative of a well defined power law behavior of the MSDC for prestellar condensations across the cloud within two orders of magnitude. The linear fit and the slope near -1 are similar to values obtained by \cite{Hartmann:2002bj} and \cite{Kraus:2008aa} for the MSDC of single stars in Taurus over a similar range of scales. 

\subsubsection{Clustering of peaks? \label{s:results:ss:mergedcores:ss:corespatialdist}}

We constructed a simple map of the surface density of extinction peaks as number of peaks per square degree, using a Nyquist sampled square grid with spatial resolution of $0.5^\circ$. The resultant map is shown in Figure \ref{fig:csd1}. From the map, we see that the Bowl is the region with the largest density of peaks per unit area, followed by B59. The Stem and the Shank regions have lower but similar total surface densities, closer to the cloud average. The Smoke region appears to be a moderately dense region in the Pipe Nebula, despite its peaks being less embedded. The map also shows that the surface density of extinction peaks is not uniform across the cloud but displays instead distinct groupings of peaks. We calculated the nearest neighbor distance between local surface density maxima and found a very narrow distribution, with a median value of $0.84\pm 0.14^\circ$, or $1.90\pm 0.32$ pc. The distribution is shown in Figure \ref{fig:csd2}. This median distance is very close to the expected Jeans length (2.0 pc) for a mean density of $10^2 \mbox{ cm}^{-3}$ and $T_K= 10$ K. Such values may be typical of the more diffuse medium traced by $^{12}\mbox{CO}$.

\subsection{Substructure or Independent Cores? \label{s:results:ss:mergedcores}}

In the analysis of the 2MASS map by RLA09, they found that about 45\% of the extinction peaks they detected were separated by less than a Jeans length. They reasoned that unless large gas motions occur in the plane of the sky, closely separated peaks with similar values of radial velocity must represent sub-structure belonging to the same physical entity, a cloud \textit{core}. Consequently they merged those extinction peaks (about 20\% of the total) which were separated by less than a Jeans length and had radial velocity differences smaller than the one-dimensional projected sound speed in a 10K gas ($\delta v<0.12$ km/s). This way they obtained a final list of 134 cores out of 158 extinction peaks. Although empirical, the prescription of RLA09 is consistent with defining a core as a bounded Jeans sized region with coherent velocity \citep{Goodman:1998aa,Bergin:2007aa}. The use of such a prescription to define a core based on a Jeans length scale is in part justified by the fact that the global properties of these cores suggest that they are a collection of thermally supported structures \citep{Lada:2008aa}. In most cases the cores have thermally dominated gas motions, as deduced from their narrow line widths (RLA09; Paper I).

In our analysis, we find that 163 out of 237 peaks are separated by their nearest neighboring peak by less than the median Jeans Length. Based on RLA09 we know that at least a fraction of them could be grouped into multiple peaked cores (for example, we used this information to merge some of the peaks of B59 in Paper I). The main result of such merging is that the mass distribution function of peaks is different from the mass distribution function of cores. Unfortunately, we do not have radial velocity information for all of the peaks we detected in our maps, so we cannot know the total fraction of peaks that could be merged into multiple peaked cores. For instance, in the study of RLA09 about half of sub-Jeans peaks were not close enough in radial velocity to belong to the same core, so we could expect this to occur for a significant number of our detections. In table \ref{tab:peaks} we used the numbering in the list of RLA09 to name those peaks separated from one of their listed positions by less than one Jeans length, using an alphabetic subindex to identify them from the most to the less massive. We compared this list against the list of RLA09 and found 93 peaks matching in position, and thus having velocity information. In most cases,
the C$^{18}$O pointing coincided with the most massive peak in a group. Only in a handful of cases a second peak in a group was also observed in C$^{18}$O and we were able to tell if those two peaks in the group were also close in velocity (we marked these cases with an asterisk). However, this information is still insufficient for the majority of our peaks and we cannot make a complete list of merged cores as in RLA09.

\subsection{Cumulative Mass Distribution \label{s:analysis:ss:cumassdist}}

As a cloud evolves toward star formation, an increasing fraction of material will be above a given threshold extinction, at which the cloud is likely to reach the critical densities for gravitational dominance. By comparing the cumulative mass distribution for different clouds with expected different levels of activity, \citet{Lombardi:2008aa} and \citet{Lada:2009ab} showed that clouds with higher levels of high formation activity (e.g. Ophiuchus vs. the Pipe Nebula and Orion vs. the California Nebula) also have a significantly larger fraction of material with high column densities. While the Pipe Nebula is known to have one of the lowest levels of star forming activity, our maps give us a good opportunity to compare the cumulative mass distribution of the five different regions of the cloud we mapped.

In Figure \ref{fig:cumul_mass} we plot the cumulative mass fraction as a function of extinction for each of the large scale maps. The curves were constructed by counting the mass of the cloud above ${A_V}=2.0$ mag. The resultant profiles for the five regions of the cloud we mapped are, indeed, very different from each other. The fraction of the mass located at high extinction varies dramatically from the outer layers in the Smoke to the cluster forming region in B59. For example, the fraction of pixels with values above $A_V=10$ mag is only over half percent in the Smoke, the Stem and the Shank regions, but although extinction rises as high as ${A_V}\approx30$ mag in the Smoke (toward B68), it never reaches a level of 20 mag in the Stem-Shank transition region. At the Bowl region the fraction of pixels with ${A_V}>10$ mag is already over 5 percent and then it is 3 times higher at B59. Therefore, in the Pipe Nebula the fraction of high density material appears to be directly related to the condensation of the material toward star formation. Our data shows that a significant increase in the fraction of high extinction material as star formation is onset, is not only noticeable from cloud to cloud, but within regions in the same cloud. Moreover, the cumulative mass fraction in the Pipe Nebula appears to vary significantly even though its global star formation rate is low.


\section{Discussion \label{s:discussion}}

One of the main goals of this paper is to take advantage of the higher spatial resolution achieved and determine the structure of the cloud across a larger range of spatial scales. From the analysis of the MSDC presented in section \ref{s:results:ss:peakspatialdist}, we see that the Pipe Nebula presents correlated dense structure between $1.45\times10^4$ AU (0.07 pc) and $1.48\times10^6$ AU (7.2 pc). The function $\Sigma_\theta$ appears to break below the former scale, indicating a significant decrease in structure of the cloud. This is consistent with the sharp decrease of the number of extinction peaks with diameters below $1.65\times 10^4$ AU ($0.08$ pc; see Fig. \ref{fig:pixelcomp}). This possible limit for the scale of fragmentation is also consistent with a recent study by \citet{Schnee:2010aa} who, using interferometric radio continuum observations, found no evidence of fragmentation in cores of the Perseus Molecular Cloud at scales of $10^3-10^4$ AU. Also, this break in the slope of the MSDC for extinction peaks appears at a scale that is smaller than the Jeans length in the cloud, $4.1\times10^4$ AU (0.2 pc). A thermal fragmentation scale is expected to be relevant in a cloud like the Pipe, where it has been found that gas motions inside cores are mostly thermal and subsonic \citep{Lada:2008aa}, so it is worth discussing some of the possible scenarios in which a significant amount of structure below the thermal fragmentation length could have formed.

One possibility is that the cloud is presenting a self-similar structure where as we increase the spatial resolution of our maps, the amount of structure also increases. We do observe a self-similar behavior within a range of spatial scales, as shown in Fig. 12. However, we observe that this behavior is truncated at scales below 0.07 pc, defining an intrinsic length in the cloud fragmentation process. The slope near -1 in the MSDC of extinction peaks in the Pipe Nebula is similar to the one obtained in the analysis of the MSDC for single stars in Taurus by \citet{Hartmann:2002bj}. They suggested that a slope near -1 could reflect the distribution of peaks along the filamentary structure of the cloud, and not necessarily a self-similar structure of the cloud.

A second possibility is that the thermal fragmentation scale is smaller than that given from equation \ref{eqn:LJL} for the median density of peaks. In order to match the Jeans length defined in equation \ref{eqn:LJL} with the observed scale of fragmentation --assuming the estimates of temperature $\sim 10$K from ammonia observations by \cite{Rathborne:2008aa}--, peaks would need to have densities about 6.3 times larger than those we derive. Likewise, if we assume the densities are correct, then the dense regions in the cloud would have to be as cold as 5 K. Both of this possibilities are unlikely. 

A third possibility is that the main filament of the Pipe Nebula is inclined by a large angle with respect to the plane of the sky, reducing the projected separations between peaks. However, a discrepancy of a factor of 2 would require  the filament to be inclined by about 60$^\circ$, which seems unlikely. However, a similar effect could be caused by the overlap of distinct regions of the cloud along the same line of sight, and this would be difficult to rule out. 

A fourth possibility is that a significant fraction of peaks separated by less than the Jeans length could belong to the same core, as found in RLA09. In that case, such peaks would represent substructure in multiple peaked cores. It could be possible that cores with multiple peaks are similar in nature to the Globule 2 in the Coalsack, a relatively massive (6 M$_\odot$) core that is characterized by a ring-like (sub-Jeans) structure \citep{Lada:2004aa} that appears to be the product of two converging subsonic flows of gas \citep{Rathborne:2009ab}, and thus in a very early stage of evolution. However
we do not have sufficient data to asses how large this effect could be across the cloud.

Another result of this study is that the analysis of the MSDC shows that individual extinction peaks are spatially distributed along the Pipe Nebula cloud in a similar fashion as single stars in the Taurus molecular cloud. At large scales, our data shows that peaks in the Pipe are distributed in small groups across the cloud, in a similar fashion to stars distributed in small clusters in Taurus (\citealt{Gomez:1993aa}). The typical separation between groups may be consistent with the Jeans length for ta more diffuse medium. Thus, our data could be evidence of thermal fragmentation of the cloud at two spatial scales, one defined by material with densities near $100\mbox{ cm}^{-3}$, and the other one defined by material with densities near $10^4\mbox{ cm}^{-3}$. The distribution of peaks is essentially similar to the distribution of stars in Taurus. Our results suggest that the spatial distribution of the stars may evolve directly from the primordial distribution of dense cloud material.

\section{Summary \label{s:summary}}
\begin{enumerate}
\item  This article describes a sensitive, high resolution, near-infrared survey of selected high extinction fields in the Pipe Nebula. The aim of the survey is to resolve the structure of individual high density regions and their local environment.

\item  The near-infrared data was used to construct extinction maps for individual fields using the NICER color excess technique. We merged our deep, high resolution photometry catalogs with data from the 2MASS survey and made five large scale maps corresponding to the main regions of the cloud: B59 (discussed separately in Paper I), the Stem, the Shank, the Bowl, and the Smoke. 

\item The new extinction maps have a spatial resolution of 20$\arcsec$, three times higher than the one achieved in the 2MASS map of LAL06, and enable us to resolve structure down to 2600 AU.  

\item We detect 244 significant extinction peaks above a more diffuse background. This number of peaks is larger than that found in the map of Lombardi et al (201 peaks). In particular, 90 peaks were found below the detection threshold of that previous survey and contribute to define a new lower completeness limit for extinction features in the Pipe Nebula. 

\item The distribution of peak to peak separations has a median value of 0.08 pc, which is coincident with a maximum value for the two-point correlation function and the mean surface density of companions near 14500 AU (0.07 pc), marking a scale below which we observe a significant decrease in the structure of the cloud.

\item The extinction peaks have masses between 0.1 and 18.4 $M_\odot$, diameters between 0.06 and 0.28 pc, and mean densities near $1.0\times$10$^4$ cm$^{-3}$, in good agreement with previous studies. The mean diameter of these extinction peaks is about half the Jeans length for this cloud. 

\item The cumulative mass fraction of pixels as a function of extinction, exhibits significant differences for separate regions in the cloud, with the cluster forming region B59 having a significantly larger fraction of material with high column densities. This shows that a significant increase in the fraction of high extinction material as star formation is onset, is not only noticeable from cloud to cloud, but within regions in the same cloud.

\item The mean surface density of companions for peaks in the Pipe Nebula is similar to the one obtained for single stars in Taurus. In addition, the surface density of the peaks is not uniform, but instead it displays several local maxima across the cloud, showing that the peaks may form small clusters, like the stars in Taurus. Our data suggests that the spatial distribution of stars we see in clouds like Taurus may evolve directly from a primordial spatial distribution of density peaks.

\end{enumerate}

\acknowledgments

We want to thank an anonymous referee for a comprehensive review of the manuscript, which resulted in a significant improvement of the content. We want to acknowledge Benoit Vandame for providing a code to produce the wavelet filtered maps, and for his very illustrative tutorials on the topic. We acknowledge fruitful discussions with Doug Johnstone (specially on unsharp masking), Jill Rathborne (who kindly shared information
on C$^{18}$O observations), August Muench, Nestor S\'anchez, Miguel Cervi\~no and Paula Teixeira, as well as discussions with the participants of the ``Pipe Nebula State of the Union'' workshop that took place in Granada in May 2009. This project acknowledges support from NASA Origins Program (NAG 13041), NASA Spitzer Program GO20119 and JPL contract 1279166. Carlos Rom\'an-Z\'u\~niga acknowledges support from a Calar Alto Post-doctoral Fellowship. Jo\~ao Alves thanks a starting research grant from the University of Vienna.  We acknowledge the help of Paranal Observatory, La Silla Observatory, and Calar Alto Observatory science operation teams for assistance during observations. Data in this publications is based on observations collected at the Centro Astron\'omico Hispano Alem\'an (CAHA), operated jointly by the Max-Planck Institut f\"ur Astronomie and the Instituto de Astrof\'isica de Andaluc\'ia (CSIC). This publication makes use of data products from the Two Micron All Sky Survey, which is a joint project of the University of Massachusetts and the Infrared Processing and Analysis Center/California Institute of Technology, funded by the National Aeronautics and Space Administration and the National Science Foundation.

{\it Facilities:}  \facility{NTT (SOFI)}, \facility{VLT:Melipal (ISAAC)}, \facility{CAHA:3.5m (OMEGA 2000)}

\clearpage

\appendix 

\section{Summary of Observations \label{s:app:observations}}

\begin{deluxetable}{llcccccl}
\tablecolumns{8}
\tablewidth{0pc}
\tablenum{A.1}
\tablecaption{Near-Infrared Observations of Pipe Nebula Fields\label{tab:obs}} 
\tablehead{
\colhead{Field ID} &
\colhead{Date Obs.} &
\multicolumn{2}{c}{Center Coords.} &
\colhead{Filter} &
\colhead{Seeing} &
\colhead{LF Peak\tablenotemark{a}} &
\colhead{Map No.\tablenotemark{b}}\\
\colhead{} &
\colhead{} &
\multicolumn{2}{c}{J2000} &
\colhead{} &
\colhead{[$(\arcsec)$]} &
\colhead{[mag]} &
\colhead{}\\
}
\startdata

\multicolumn{8}{c}{SOFI-NTT OBSERVATIONS}\\*
\cline{1-8}\\*	

 B59-01    &   2002-06-20   &  257.631573  &   -27.426578  & 	$H$   &   0.86  &  19.25 & A.1, 3 \\		  
 B59-01    &   2002-06-20   &  257.626906  &   -27.428854  &   $K_s$  &   0.81  &  18.50 & A.1, 3 \\ 
 B59-02    &   2002-06-20   &  257.722044  &   -27.431189  & 	$H$   &   1.00  &  19.25 & A.2, 3 \\ 
 B59-02    &   2002-06-20   &  257.723116  &   -27.431726  &   $K_s$  &   0.93  &  18.25 & A.2, 3 \\ 
 B59-03    &   2002-06-20   &  257.721593  &   -27.459140  & 	$H$   &   0.90  &  18.75 & A.3, 3 \\ 
 B59-03    &   2002-06-20   &  257.719551  &   -27.460034  &   $K_s$  &   0.85  &  17.75 & A.3, 3 \\ 
 B59-04    &   2001-06-09   &  257.934286  &   -27.432649  & 	$H$   &   1.06  &  18.75 &  3 \\ 
 B59-04    &   2001-06-09   &  257.935579  &   -27.433704  &   $K_s$  &   1.09  &  18.25 &  3 \\ 
 B59-05    &   2002-06-20   &  257.799520  &   -27.470299  & 	$H$   &   0.84  &  19.75 &  3 \\ 
 B59-05    &   2002-06-20   &  257.803101  &   -27.472615  &   $K_s$  &   0.84  &  18.75 &  3 \\  
 B59-06    &   2002-06-20   &  257.914017  &   -27.374625  & 	$H$   &   0.89  &  19.75 &  3 \\ 
 B59-06    &   2002-06-20   &  257.914131  &   -27.375762  &   $K_s$  &   0.83  &  18.75 &  3 \\ 
 B59-07    &   2002-06-20   &  257.910182  &   -27.469140  & 	$H$   &   0.80  &  19.75 &  3 \\ 
 B59-07    &   2002-06-20   &  257.910421  &   -27.467422  &   $K_s$  &   0.78  &  18.25 &  3 \\ 
 B59-08    &   2002-06-20   &  257.909318  &   -27.550688  & 	$H$   &   0.85  &  19.25 & A.4, 3 \\ 
 B59-08    &   2002-06-20   &  257.916099  &   -27.550724  &   $K_s$  &   0.76  &  18.75 & A.4, 3 \\ 
 B59-09    &   2002-06-20   &  258.066477  &   -27.621523  & 	$H$   &   0.95  &  18.75 & A.5, 3 \\ 
 B59-09    &   2002-06-20   &  258.068466  &   -27.623340  &   $K_s$  &   0.84  &  18.25 & A.5, 3 \\ 
 B59-10    &   2002-06-20   &  257.986905  &   -27.426153  & 	$H$   &   0.85  &  19.75 & A.6, 3 \\ 
 B59-10    &   2002-06-20   &  257.987585  &   -27.423171  &   $K_s$  &   0.79  &  18.25 & A.6, 3 \\ 
 B59-11    &   2002-06-20   &  258.073628  &   -27.410908  & 	$H$   &   0.84  &  19.25 & A.6, 3 \\ 
 B59-11    &   2002-06-20   &  258.073722  &   -27.411663  &   $K_s$  &   0.82  &  18.25 & A.6, 3 \\ 
 B59-12    &   2002-06-20   &  258.138957  &   -27.340032  & 	$H$   &   0.84  &  19.25 & A.6, 3 \\ 
 B59-12    &   2002-06-20   &  258.139848  &   -27.339383  &   $K_s$  &   0.79  &  18.25 & A.6, 3 \\ 
 B59-13    &   2002-06-21   &  258.220558  &   -27.391059  & 	$H$   &   0.97  &  18.75 & A.7, 3 \\ 
 B59-13    &   2002-06-21   &  258.219609  &   -27.392223  &   $K_s$  &   0.91  &  18.25 & A.7, 3 \\
Pipe-16    &   2002-06-22   &  258.815340  &   -27.557691  & 	$H$   &   0.88  &  18.25 & A.8, 4 \\
Pipe-16    &   2002-06-22   &  258.815775  &   -27.558311  &   $K_s$  &   0.77  &  17.75 & A.8, 4 \\
Pipe-17    &   2002-06-22   &  258.943817  &   -27.503255  & 	$H$   &   0.98  &  18.75 & A.9, 4 \\
Pipe-17    &   2002-06-22   &  258.942134  &   -27.502738  &   $K_s$  &   0.81  &  17.75 & A.9, 4 \\
Pipe-18    &   2002-06-22   &  259.021660  &   -27.517372  & 	$H$   &   0.84  &  18.25 & A.9, 4 \\
Pipe-18    &   2002-06-22   &  259.020609  &   -27.517635  &   $K_s$  &   0.78  &  17.75 & A.9, 4 \\
Pipe-19    &   2002-06-22   &  258.743043  &   -27.368822  & 	$H$   &   0.87  &  18.25 & A.10, 4 \\
Pipe-19    &   2002-06-22   &  258.742166  &   -27.369409  &   $K_s$  &   0.88  &  17.75 & A.10, 4 \\
Pipe-20    &   2002-06-22   &  258.726571  &   -27.291624  & 	$H$   &   0.74  &  18.25 & A.10, 4 \\
Pipe-20    &   2002-06-22   &  258.729465  &   -27.290965  &   $K_s$  &   0.72  &  17.25 & A.10, 4 \\
Pipe-21    &   2002-06-22   &  259.096244  &   -27.169525  & 	$H$   &   0.75  &  18.75 & A.11, 4 \\
Pipe-21    &   2002-06-22   &  259.096514  &   -27.167508  &   $K_s$  &   0.69  &  17.75 & A.11, 4 \\
Pipe-22A   &   2002-06-22   &  259.291270  &   -27.029664  & 	$H$   &   0.73  &  18.75 & A.12, 4 \\
Pipe-22A   &   2002-06-22   &  259.291548  &   -27.031440  &   $K_s$  &   0.70  &  17.75 & A.12, 4 \\
Pipe-22B   &   2002-06-22   &  259.347697  &   -27.116944  & 	$H$   &   0.66  &  18.75 & A.12, 4 \\
Pipe-22B   &   2002-06-22   &  259.346677  &   -27.117245  &   $K_s$  &   0.66  &  17.75 & A.12, 4 \\
Pipe-23    &   2002-06-22   &  259.627722  &   -26.806936  & 	$H$   &   0.68  &  18.75 & A.13, 4 \\
Pipe-23    &   2002-06-22   &  259.629572  &   -26.805582  &   $K_s$  &   0.64  &  17.75 & A.13, 4 \\
Pipe-24    &   2002-06-22   &  259.897755  &   -26.724130  & 	$H$   &   0.71  &  18.75 & A.14, 4 \\
Pipe-24    &   2002-06-22   &  259.896565  &   -26.721405  &   $K_s$  &   0.67  &  17.75 & A.14, 4 \\
Pipe-25    &   2002-06-21   &  259.922919  &   -26.924565  & 	$H$   &   0.89  &  18.75 & A.15, 4 \\
Pipe-25    &   2002-06-21   &  259.921073  &   -26.928810  &   $K_s$  &   0.85  &  18.25 & A.15, 4 \\
Pipe-26    &   2002-06-22   &  260.063937  &   -26.997820  & 	$H$   &   0.90  &  18.75 & A.16, 4 \\
Pipe-26    &   2002-06-22   &  260.062432  &   -26.998108  &   $K_s$  &   0.90  &  17.75 & A.16, 4 \\
Pipe-27    &   2002-06-22   &  260.243891  &   -26.806044  & 	$H$   &   1.11  &  18.75 & A.17, 4 \\
Pipe-27    &   2002-06-22   &  260.245120  &   -26.807443  &   $K_s$  &   0.97  &  17.75 & A.17, 4 \\
Pipe-28    &   2002-06-21   &  260.244531  &   -26.889373  & 	$H$   &   0.80  &  18.25 & A.17, 4 \\
Pipe-28    &   2002-06-21   &  260.243043  &   -26.888645  &   $K_s$  &   0.77  &  17.75 & A.17, 4 \\
Pipe-29    &   2002-06-21   &  260.344398  &   -26.885212  & 	$H$   &   0.83  &  18.75 & A.17, 4 \\
Pipe-29    &   2002-06-21   &  260.343846  &   -26.886681  &   $K_s$  &   0.79  &  17.75 & A.17, 4 \\
Pipe-30    &   2002-06-21   &  260.609264  &   -27.075281  & 	$H$   &   0.81  &  18.75 & A.18, 4 \\
Pipe-30    &   2002-06-21   &  260.608431  &   -27.074964  &   $K_s$  &   0.81  &  17.75 & A.18, 4 \\
Pipe-31    &   2002-06-21   &  260.700080  &   -27.072468  & 	$H$   &   0.76  &  18.75 & A.18, 4 \\
Pipe-31    &   2002-06-21   &  260.699113  &   -27.072382  &   $K_s$  &   0.78  &  17.75 & A.18, 4 \\
Pipe-32    &   2002-06-22   &  261.487728  &   -26.746283  & 	$H$   &   0.89  &  18.75 & A.19, 5 \\
Pipe-32    &   2002-06-22   &  261.487492  &   -26.744937  &   $K_s$  &   1.10  &  17.75 & A.19, 5 \\
Pipe-33    &   2002-06-22   &  261.840106  &   -26.740218  & 	$H$   &   0.86  &  18.75 & A.20, 5 \\
Pipe-33    &   2002-06-22   &  261.840850  &   -26.739058  &   $K_s$  &   0.76  &  17.75 & A.20, 5 \\
Pipe-34    &   2002-06-22   &  261.855804  &   -26.974552  & 	$H$   &   0.87  &  17.75 & A.21, 5 \\
Pipe-34    &   2002-06-22   &  261.854675  &   -26.974129  &   $K_s$  &   0.82  &  17.75 & A.21, 5 \\
Pipe-35    &   2000-03-13   &  262.041197  &   -26.378789  & 	$H$   &   0.87  &  18.75 & A.22, 5 \\
Pipe-35    &   2000-03-13   &  262.041080  &   -26.379079  &   $K_s$  &   0.82  &  17.75 & A.22, 5 \\
Pipe-36    &   2000-03-13   &  262.040733  &   -26.460077  & 	$H$   &   0.91  &  17.25 & A.22, 5 \\
Pipe-36    &   2000-03-13   &  262.041125  &   -26.460572  &   $K_s$  &   1.02  &  17.25 & A.22, 5 \\
Pipe-37    &   2002-06-22   &  262.199115  &   -26.308098  & 	$H$   &   0.84  &  17.75 & A.23, 5 \\
Pipe-37    &   2002-06-22   &  262.198830  &   -26.309243  &   $K_s$  &   0.83  &  17.75 & A.23, 5 \\
Pipe-38    &   2002-06-22   &  262.394769  &   -25.906708  & 	$H$   &   0.87  &  18.25 & A.24, 5 \\
Pipe-38    &   2002-06-22   &  262.396783  &   -25.909761  &   $K_s$  &   0.80  &  17.25 & A.24, 5 \\
Pipe-39    &   2002-06-22   &  262.799700  &   -26.480744  & 	$H$   &   0.86  &  18.25 & A.25, 5 \\
Pipe-39    &   2002-06-22   &  262.802144  &   -26.480889  &   $K_s$  &   0.79  &  17.75 & A.25, 5 \\
Pipe-40    &   2002-06-22   &  262.858125  &   -26.521756  & 	$H$   &   0.84  &  18.25 & A.25, 5 \\
Pipe-40    &   2002-06-22   &  262.858611  &   -26.521687  &   $K_s$  &   0.84  &  17.25 & A.25, 5 \\
Pipe-41    &   2002-06-21   &  263.158008  &   -26.256971  & 	$H$   &   1.00  &  18.25 & A.26, 5 \\
Pipe-41    &   2002-06-21   &  263.156776  &   -26.257098  &   $K_s$  &   0.89  &  17.25 & A.26, 5 \\
Pipe-42    &   2002-06-22   &  263.001816  &   -25.422071  & 	$H$   &   0.85  &  18.25 & A.27, 6 \\
Pipe-42    &   2002-06-22   &  263.001133  &   -25.421451  &   $K_s$  &   0.84  &  17.75 & A.27, 6 \\
Pipe-43    &   2002-06-22   &  263.081523  &   -25.420874  & 	$H$   &   0.80  &  17.75 & A.27, 6 \\
Pipe-43    &   2002-06-22   &  263.080633  &   -25.418784  &   $K_s$  &   0.80  &  17.25 & A.27, 6 \\
Pipe-44    &   2002-06-21   &  263.524876  &   -25.838765  & 	$H$   &   0.88  &  17.75 & A.28, 6 \\
Pipe-44    &   2002-06-21   &  263.523630  &   -25.838433  &   $K_s$  &   0.81  &  17.25 & A.28, 6 \\
Pipe-45    &   2002-06-21   &  263.614868  &   -25.839539  & 	$H$   &   0.84  &  18.75 & A.28, 6 \\
Pipe-45    &   2002-06-21   &  263.612704  &   -25.839888  &   $K_s$  &   0.83  &  17.75 & A.28, 6 \\
Pipe-46    &   2002-06-21   &  263.683934  &   -25.790023  & 	$H$   &   0.90  &  18.75 & A.28, 6 \\
Pipe-46    &   2002-06-21   &  263.685809  &   -25.791936  &   $K_s$  &   0.83  &  17.75 & A.28, 6 \\
Pipe-47    &   2002-06-21   &  263.587565  &   -25.568535  & 	$H$   &   0.77  &  18.25 & A.29, 6 \\
Pipe-47    &   2002-06-21   &  263.588266  &   -25.564001  &   $K_s$  &   0.79  &  17.75 & A.29, 6 \\
Pipe-48    &   2002-06-21   &  263.384176  &   -25.509374  & 	$H$   &   0.85  &  18.75 & A.30, 6 \\
Pipe-48    &   2002-06-21   &  263.382664  &   -25.509442  &   $K_s$  &   0.73  &  17.75 & A.30, 6 \\
Pipe-49    &   2002-06-21   &  263.366966  &   -25.684444  & 	$H$   &   0.90  &  18.25 & A.31, 6 \\
Pipe-49    &   2002-06-21   &  263.369715  &   -25.683947  &   $K_s$  &   0.81  &  17.75 & A.31, 6 \\
Pipe-50    &   2002-06-21   &  263.510308  &   -25.665333  & 	$H$   &   0.90  &  18.25 & A.31, 6 \\
Pipe-50    &   2002-06-21   &  263.511060  &   -25.663561  &   $K_s$  &   0.86  &  17.75 & A.31, 6 \\
Pipe-51    &   2002-06-22   &  263.451184  &   -25.740991  & 	$H$   &   0.78  &  18.75 & A.31, 6 \\
Pipe-51    &   2002-06-22   &  263.451698  &   -25.741033  &   $K_s$  &   0.77  &  17.75 & A.31, 6 \\
Pipe-52A   &   2002-06-22   &  263.288701  &   -25.382022  & 	$H$   &   0.81  &  18.25 & A.32, 6 \\
Pipe-52A   &   2002-06-22   &  263.289384  &   -25.382309  &   $K_s$  &   0.80  &  17.75 & A.32, 6 \\
Pipe-53    &   2002-06-22   &  263.556590  &   -25.486642  & 	$H$   &   0.79  &  18.75 & A.29, 6 \\
Pipe-53    &   2002-06-22   &  263.556590  &   -25.486642  &   $K_s$  &   0.81  &  17.75 & A.20, 6 \\
Pipe-54    &   2002-06-22   &  263.537640  &   -25.414060  & 	$H$   &   0.89  &  18.25 & A.29, 6 \\
Pipe-54    &   2002-06-22   &  263.537640  &   -25.414060  &   $K_s$  &   0.79  &  17.75 & A.29, 6 \\
Fest-1457  &   2000-03-13   &  263.946902  &   -25.556839  &   $H$    &	  0.78  &  17.75 & A.33, 6 \\
Fest-1457  &   2000-03-13   &  263.948758  &   -25.557231  &   $J$    &	  0.81  &  19.25 & A.33, 6 \\
Fest-1457  &   2000-03-13   &  263.948785  &   -25.556266  &   $K_s$  &   0.71  &  17.75 & A.33, 6 \\
Pipe-75    &   2002-06-22   &  261.270282  &   -24.204082  &    $H$   &	  1.30  &  17.25 & A.34, 7 \\
Pipe-75    &   2002-06-22   &  261.266556  &   -24.205522  &   $K_s$  &   1.12  &  16.75 & A.34, 7 \\
Pipe-77    &   2002-06-22   &  260.903473  &   -24.047282  &   $H$    &	  1.22  &  17.75 & A.35, 7 \\
Pipe-77    &   2002-06-22   &  260.903205  &   -24.048410  &   $K_s$  &   1.21  &  17.25 & A.35, 7 \\
B68-01     &   2001-03-07   &  260.663570  &   -23.831128  &    $H$   &	  0.80  &  18.25 & A.36, 7 \\
B68-01     &   2000-03-14   &  260.663878  &   -23.830876  &   $K_s$  &   0.80  &  18.25 & A.36, 7 \\
B72E-01    &   2000-03-13   &  260.974581  &   -23.713071  &    $H$   &	  0.87  &  18.25 & A.37, 7 \\
B72E-01    &   2000-03-13   &  260.974438  &   -23.713214  &   $K_s$  &   0.93  &  17.75 & A.37, 7 \\
B72W-01    &   2000-03-13   &  260.894930  &   -23.675185  &    $H$   &	  0.92  &  18.25 & A.37, 7 \\
B72W-01    &   2000-03-13   &  260.893699  &   -23.675714  &   $K_s$  &   0.74  &  18.25 & A.37, 7 \\
Pipe-00    &   2002-06-22   &  257.043428  &   -28.050660  &    $H$   &	  0.87  &  17.75 & control field \\
Pipe-00    &   2002-06-22   &  257.045819  &   -28.052072  &    $J$   &	  0.91  &  19.25 & control field \\
Pipe-00    &   2002-06-22   &  257.042540  &   -28.052401  &   $K_s$  &   0.79  &  18.25 & control field \\

\cline{1-8}\\*	
\multicolumn{8}{c}{ISAAC-VLT OBSERVATIONS}\\*
\cline{1-8}\\*	

B59-1i     &   2002-07-26   &  257.630470  &   -27.424219  &    $H$   & 0.51  &  24.50 & 3 \\
B59-1i     &   2002-07-26   &  257.629403  &   -27.426465  &   $K_s$  & 0.53  &  24.00 & 3 \\
B59C-A     &   2002-07-27   &  257.857306  &   -27.411161  &    $H$   & 0.57  &  24.50 & 3 \\ 
B59C-A     &   2002-07-27   &  257.857584  &   -27.411213  &   $K_s$  & 0.56  &  24.00 & 3 \\
B59C-B     &   2002-07-29   &  257.829390  &   -27.448323  &    $H$   & 0.74  &  24.00 & 3 \\
B59C-B     &   2002-07-29   &  257.829693  &   -27.448552  &   $K_s$  & 0.90  &  23.75 & 3 \\
Pipe-29i   &   2002-07-28   &  260.314081  &   -26.888345  &    $H$   &	0.57  &  20.25 & A.17, 4 \\
Pipe-29i   &   2002-07-28   &  260.314563  &   -26.888015  &   $K_s$  & 0.57  &  19.75 & A.17, 4 \\
Pipe-31i   &   2002-07-27   &  260.670952  &   -27.088082  &    $H$   &	0.69  &  20.25 & A.18, 4 \\
Pipe-31i   &   2002-07-27   &  260.671709  &   -27.088022  &   $K_s$  & 0.65  &  19.75 & A.18, 4 \\
Pipe-45i   &   2002-07-30   &  263.611248  &   -25.832594  &    $H$   &	0.50  &  20.25 & A.28, 6 \\
Pipe-45i   &   2002-07-30   &  263.612185  &   -25.831998  &   $K_s$  & 0.48  &  20.00 & A.28, 6 \\
Fest-1457i &   2000-06-09   &  263.949504  &   -25.557491  &	$H$   &	0.57  &  20.25 & A.33, 6 \\
Fest-1457i &   2000-06-09   &  263.950490  &   -25.555175  &   $K_s$  & 0.61  &  19.75 & A.33, 6 \\   

\cline{1-8}\\*	
\multicolumn{8}{c}{OMEGA 2000-CAHA 3.5m OBSERVATIONS}\\*
\cline{1-8}\\*	

CAHA-F01   &   2007-06-04   &  264.511300  &   -25.257382  &    $H$   &	1.53  &  16.75 & A.38, 6 \\      
CAHA-F01   &   2007-06-04   &  264.513106  &   -25.251004  &    $J$   &	1.52  &  18.25 & A.38, 6 \\      
CAHA-F01   &   2007-06-04   &  264.519161  &   -25.256174  &   $K_s$  & 1.40  &  16.75 & A.38, 6 \\      
CAHA-F02   &   2007-06-05   &  263.714344  &   -25.712625  &    $H$   &	1.15  &  17.75 & A.39, 6 \\      
CAHA-F02   &   2007-06-05   &  263.720709  &   -25.715359  &    $J$   &	1.21  &  18.75 & A.39, 6 \\      
CAHA-F02   &   2007-06-05   &  263.716068  &   -25.717830  &   $K_s$  & 1.19  &  16.75 & A.39, 6 \\      
CAHA-F03   &   2007-06-05   &  263.603253  &   -25.874419  &    $H$   &	1.10  &  17.75 & A.40, 6 \\      
CAHA-F03   &   2007-06-05   &  263.603969  &   -25.874788  &    $J$   &	1.17  &  18.75 & A.40, 6 \\      
CAHA-F03   &   2007-06-05   &  263.606448  &   -25.879535  &   $K_s$  & 1.25  &  16.75 & A.40, 6 \\      
CAHA-F04   &   2007-06-05   &  263.464723  &   -25.712945  &    $H$   &	1.13  &  17.75 & A.41, 6 \\      
CAHA-F04   &   2007-06-06   &  263.465829  &   -25.712984  &    $J$   &	1.15  &  19.25 & A.41, 6 \\      
CAHA-F04   &   2007-06-05   &  263.465142  &   -25.714312  &   $K_s$  & 1.02  &  17.25 & A.41, 6 \\      
CAHA-F05   &   2007-06-06   &  263.345584  &   -25.974453  &    $H$   &	1.12  &  17.75 & A.42, 6 \\      
CAHA-F05   &   2007-06-06   &  263.348588  &   -25.972739  &    $J$   &	1.13  &  18.75 & A.42, 6 \\      
CAHA-F05   &   2007-06-06   &  263.345820  &   -25.973940  &   $K_s$  & 1.09  &  17.25 & A.42, 6 \\      
CAHA-F06   &   2007-06-06   &  264.102635  &   -25.377285  &    $H$   &	1.11  &  17.25 & A.43, 6 \\      
CAHA-F06   &   2007-06-06   &  264.101441  &   -25.374326  &   $K_s$  & 1.34  &  16.75 & A.43, 6 \\      
CAHA-F07   &   2007-06-06   &  263.854754  &   -25.394960  &    $H$   &	1.22  &  17.25 & A.44, 6 \\      
CAHA-F07   &   2007-06-07   &  263.858910  &   -25.395218  &    $J$   &	1.30  &  18.75 & A.44, 6 \\      
CAHA-F07   &   2007-06-07   &  263.849694  &   -25.396181  &   $K_s$  & 1.54  &  16.75 & A.44, 6 \\      
CAHA-F08   &   2007-06-07   &  263.230680  &   -25.677287  &    $H$   &	1.42  &  17.25 & A.45, 6 \\      
CAHA-F08   &   2007-06-08   &  263.231877  &   -25.676369  &    $J$   &	1.44  &  18.75 & A.45, 6 \\      
CAHA-F08   &   2007-06-08   &  263.233453  &   -25.678540  &   $K_s$  & 1.16  &  17.25 & A.45, 6 \\      
CAHA-F09   &   2007-06-08   &  263.281068  &   -25.463787  &    $H$   &	1.48  &  17.75 & A.46, 6 \\      
CAHA-F09   &   2007-06-08   &  263.280621  &   -25.461972  &    $J$   &	1.36  &  18.75 & A.46, 6 \\      
CAHA-F09   &   2007-06-08   &  263.282406  &   -25.464273  &   $K_s$  & 1.24  &  17.25 & A.46, 6 \\      
CAHA-F10   &   2007-06-27   &  263.759558  &   -24.435373  &    $H$   &	1.30  &  17.25 & A.56\\      
CAHA-F10   &   2007-06-27   &  263.757858  &   -24.434254  &    $J$   &	1.30  &  18.75 & A.56\\      
CAHA-F10   &   2007-06-27   &  263.758513  &   -24.435108  &   $K_s$  & 1.11  &  17.25 & A.56\\      
CAHA-F11   &   2007-06-27   &  264.829458  &   -25.078098  &    $H$   &	1.15  &  17.25 & A.47, 6 \\      
CAHA-F11   &   2007-06-27   &  264.830654  &   -25.078716  &    $J$   &	1.16  &  18.25 & A.47, 6 \\      
CAHA-F11   &   2007-06-27   &  264.829789  &   -25.082167  &   $K_s$  & 1.20  &  16.75 & A.47, 6 \\      
CAHA-F12   &   2007-06-28   &  257.678253  &   -27.367815  &    $H$   &	1.48  &  17.75 & --- \\      
CAHA-F12   &   2007-06-28   &  257.682254  &   -27.367639  &   $K_s$  & 1.44  &  17.25 & --- \\      
CAHA-F13   &   2007-06-29   &  262.828405  &   -26.563106  &    $H$   &	1.16  &  17.75 & A.48, 6 \\      
CAHA-F13   &   2007-06-29   &  262.830957  &   -26.561892  &    $J$   &	1.14  &  18.25 & A.48, 6 \\      
CAHA-F13   &   2007-06-29   &  262.827677  &   -26.563795  &   $K_s$  & 1.11  &  17.25 & A.48, 6 \\      
CAHA-F14   &   2007-06-29   &  262.122621  &   -26.372488  &    $H$   &	1.23  &  17.75 & A.49, 6 \\      
CAHA-F14   &   2007-06-29   &  262.122746  &   -26.371671  &    $J$   &	1.20  &  18.75 & A.49, 6 \\      
CAHA-F14   &   2007-06-29   &  262.123242  &   -26.372642  &   $K_s$  & 1.14  &  17.25 & A.49, 6 \\      
CAHA-F15   &   2007-06-30   &  262.310347  &   -25.907021  &    $H$   &	1.37  &  17.75 & A.50, 6 \\      
CAHA-F15   &   2007-06-30   &  262.307427  &   -25.903850  &    $J$   &	1.47  &  18.25 & A.50, 6 \\      
CAHA-F15   &   2007-06-30   &  262.307661  &   -25.905993  &   $K_s$  & 1.44  &  17.25 & A.50, 6 \\      
CAHA-F16   &   2007-07-01   &  264.412924  &   -23.462756  &    $H$   &	1.45  &  17.25 & A.57\\      
CAHA-F16   &   2007-07-01   &  264.416402  &   -23.465940  &    $J$   &	1.60  &  18.25 & A.57\\      
CAHA-F16   &   2007-07-01   &  264.417416  &   -23.466513  &   $K_s$  & 1.34  &  16.75 & A.57\\      
CAHA-F17   &   2008-05-21   &  262.605850  &   -25.972400  &    $H$   &	1.21  &  17.75 & A.51, 6 \\	
CAHA-F17   &   2008-05-21   &  262.607893  &   -25.977116  &    $J$   &	1.31  &  18.25 & A.51, 6 \\	
CAHA-F17   &   2008-05-21   &  262.603189  &   -25.978994  &   $K_s$  & 1.27  &  16.75 & A.51, 6 \\	
CAHA-F18   &   2008-05-21   &  261.368754  &   -24.180606  &    $H$   &	1.38  &  18.25 & A.52, 7 \\	
CAHA-F18   &   2008-05-21   &  261.365402  &   -24.186541  &    $J$   &	1.20  &  18.75 & A.52, 7 \\	
CAHA-F18   &   2008-05-21   &  261.373829  &   -24.180924  &   $K_s$  & 1.39  &  17.25 & A.52, 7 \\	
CAHA-F23   &   2008-06-18   &  261.309721  &   -22.447263  &    $H$   &	1.72  &  17.75 & A.53, 7 \\	
CAHA-F23   &   2008-06-18   &  261.310504  &   -22.443345  &    $J$   &	1.82  &  18.75 & A.53, 7 \\	
CAHA-F23   &   2008-06-19   &  261.310456  &   -22.448386  &   $K_s$  & 1.74  &  17.75 & A.53, 7 \\	
CAHA-F25   &   2008-06-21   &  262.683720  &   -26.817767  &    $H$   &	1.26  &  17.25 & A.54, 6 \\	
CAHA-F25   &   2008-06-21   &  262.684713  &   -26.817338  &    $J$   &	1.34  &  18.25 & A.54, 6 \\	
CAHA-F25   &   2008-06-21   &  262.681433  &   -26.815974  &   $K_s$  & 1.41  &  16.75 & A.54, 6 \\	
CAHA-F28   &   2008-06-19   &  261.709733  &   -26.781448  &    $H$   &	1.33  &  17.75 & A.55, 6 \\	
CAHA-F28   &   2008-06-19   &  261.710778  &   -26.779476  &    $J$   &	1.39  &  18.25 & A.55, 6 \\	
CAHA-F28   &   2008-06-20   &  261.708038  &   -26.782866  &   $K_s$  & 1.30  &  17.25 & A.55, 6 \\
\enddata
\tablenotetext{a}{Expresses the value at the peak of the observed magnitude distribution} 
\tablenotetext{b}{Indicates corresponding fig. number in large scale maps (figs. 3-7) and Atlas (ONLINE MATERIAL)} 

\end{deluxetable}

\clearpage

\section{Completeness \label{s:app:completeness}}

\begin{figure}[ht]
\figurenum{B.01}
\centering
\includegraphics[width=3.0in]{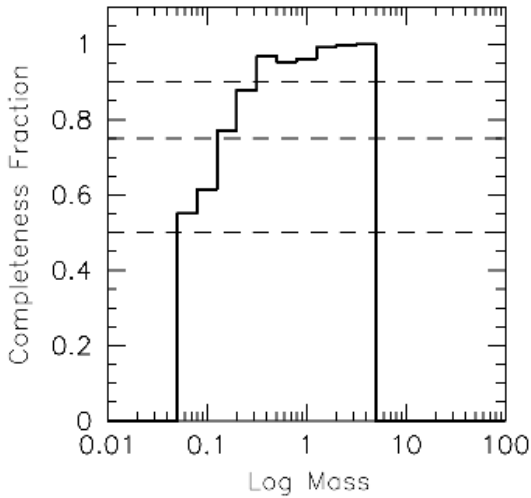}
\caption{Results for the mass completeness experiments. A successful retrieval of a simulated core is defined if a) the core is detected and b) if its CLF2D mass is within 20\% of the input mass The histogram shows the recovery fractions in the $\log{\mbox{mass}}$ space, with the dashed horizontal lines marking levels of 90, 75 and 50 percent \label{fig:completeness}}
\end{figure}

\begin{figure*}[ht]
\centering
\includegraphics[width=4.0in]{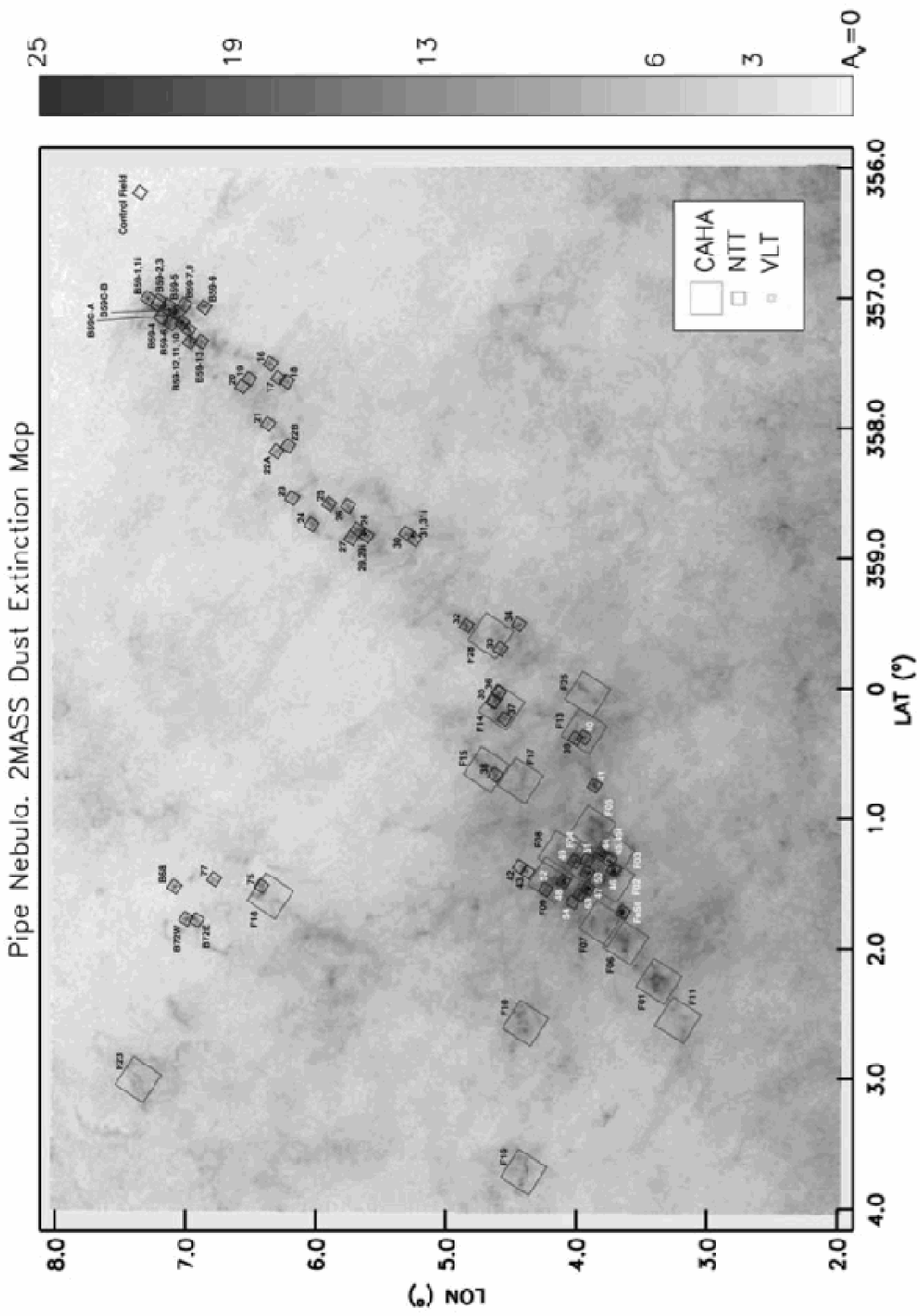}
\caption{The large field extinction map of \citet{Lombardi:2001aa} with an overlay of 
the observed fields in the high resolution near-infrared survey. Fields indicate central position and 
corresponding imager field of view. Red boxes indicate Omega-2000/CAHA-3.5m
fields, blue boxes indicate SOFI/ESO-NTT fields and green boxes indicate ISAAC/ESO-VLT fields. Labels
follow the nomenclature used in table \ref{tab:obs}. \label{fig:FOV_MAP}}
\end{figure*}

\begin{figure*}[ht]
\centering
\includegraphics[width=4.0in]{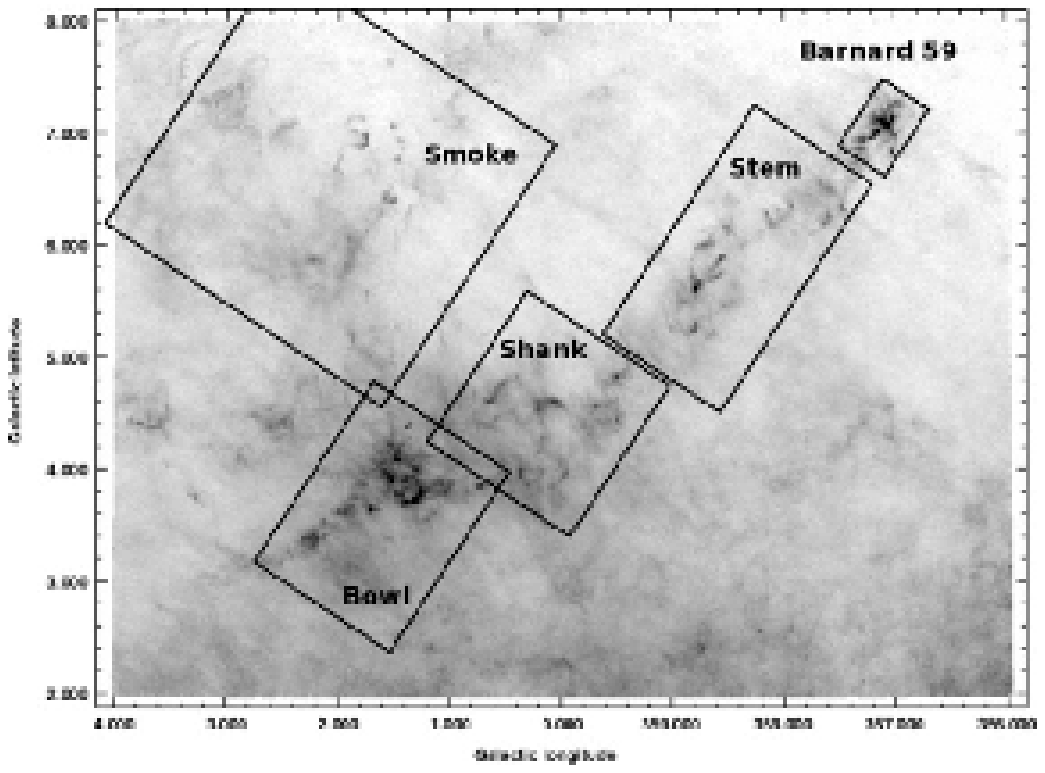}
\caption{The large field extinction map of \citet{Lombardi:2001aa} showing the
regions selected to construct hybrid maps. The extent 
of the regions is listed in table \ref{tab:mapareas}. \label{fig:MAP_REGS}}
\end{figure*}

\begin{figure*}[ht]
\centering
\includegraphics[width=7.0in]{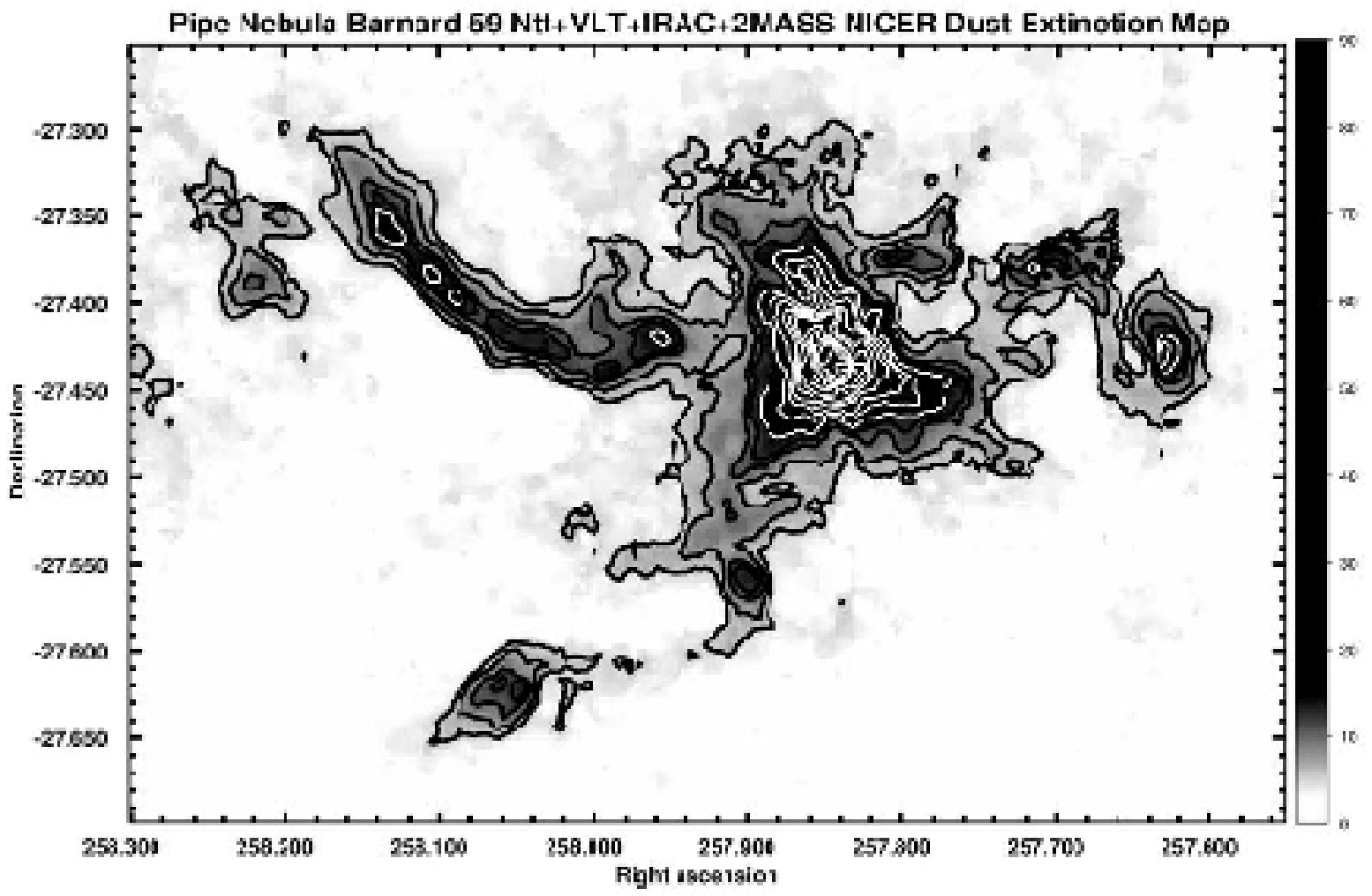}
\caption{Dust extinction map of the Barnard 59 region at a spatial resolution of 20$\arcsec$. 
The white solid line contours mark levels of visual extinction at $A_V=5,7,10,15,20,25,30,35,40,
50,60,70,80\mathrm{\ and\ }90$ mag. \textit{Please see the electronic version of the Journal for a color version of this figure.} \label{fig:LMAP_B59}}
\end{figure*}

\begin{figure*}[ht]
\centering
\includegraphics[width=7.0in]{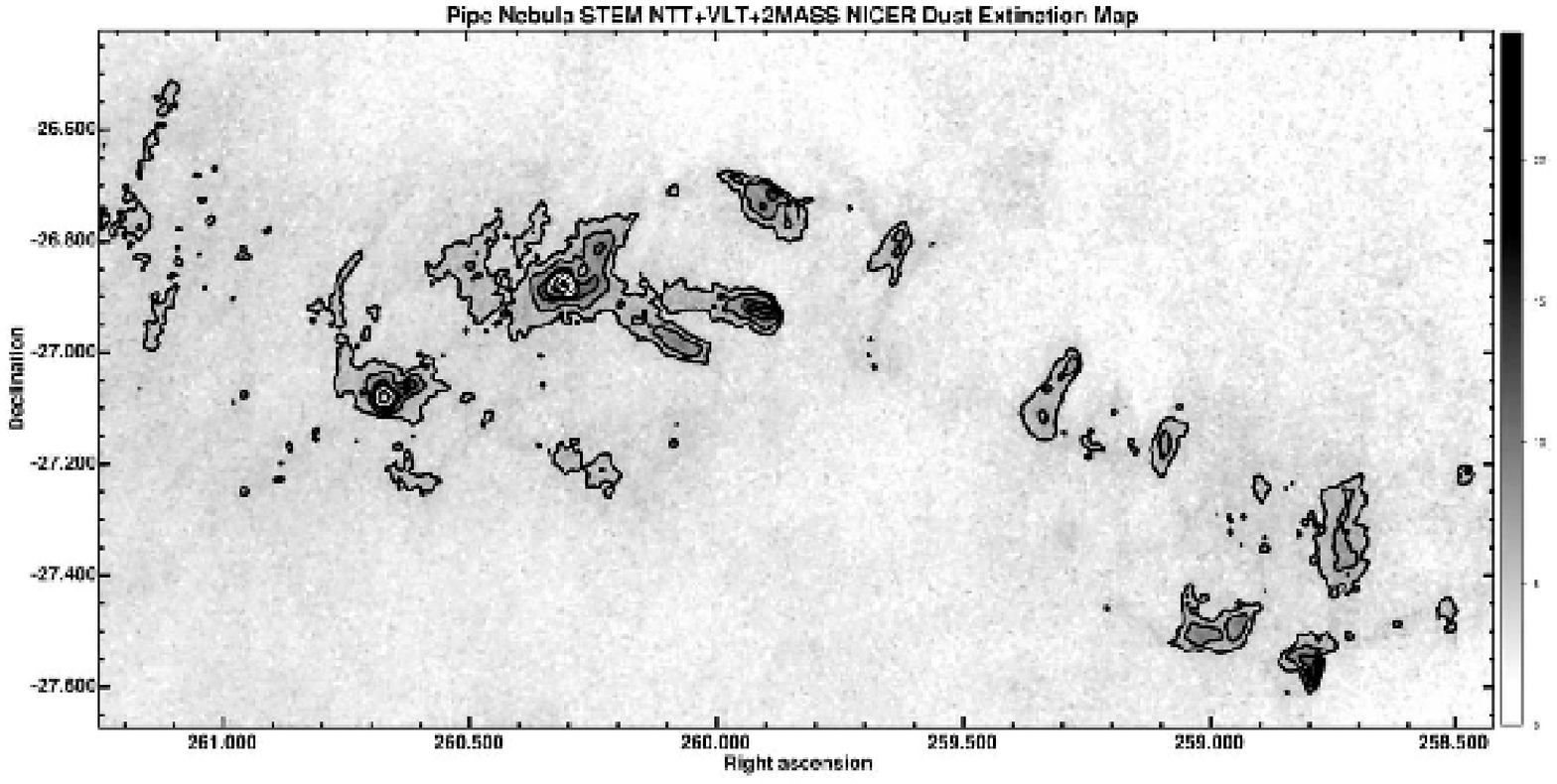}
\caption{Extinction map of the Pipe Stem region at a spatial resolution of 20$\arcsec$.
The white solid line contours mark levels of visual extinction at $A_V=3,5,7,10,12,15
\mathrm{\ and\ }20$ mag. \textit{Please see the electronic version of the Journal for a color version of this figure.}\label{fig:LMAP_STEM}}
\end{figure*}

\begin{figure*}[ht]
\centering
\includegraphics[width=7.0in]{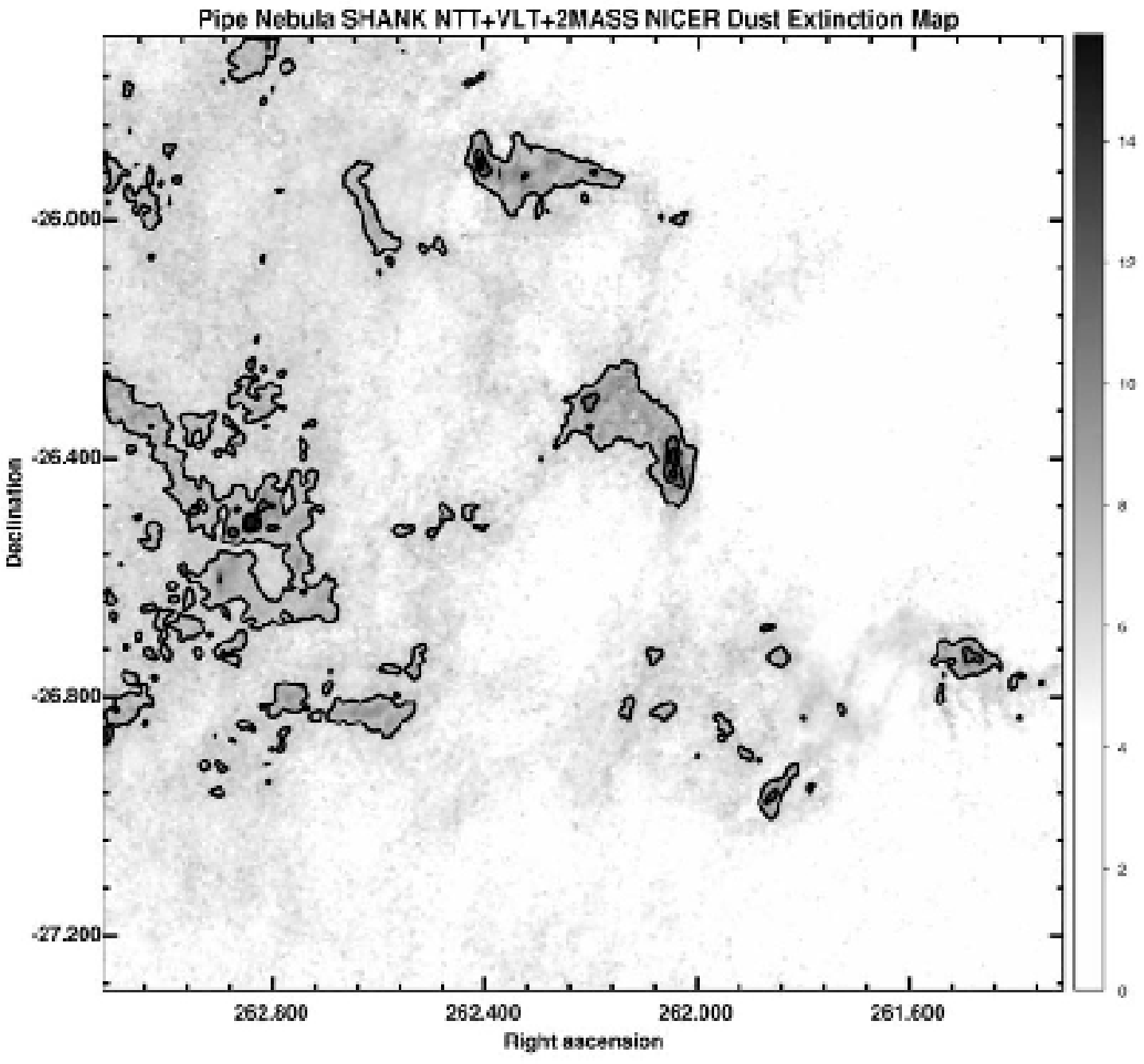}
\caption{Extinction map of the Pipe Shank region at a spatial resolution of 20$\arcsec$. 
The white solid line contours mark levels of visual extinction at $A_V=3,5,7,10,12
\mathrm{\ and\ }15$ mag. \textit{Please see the electronic version of the Journal for a color version of this figure.} \label{fig:LMAP_SHANK}}
\end{figure*}

\begin{figure*}[ht]
\centering
\includegraphics[width=7.0in]{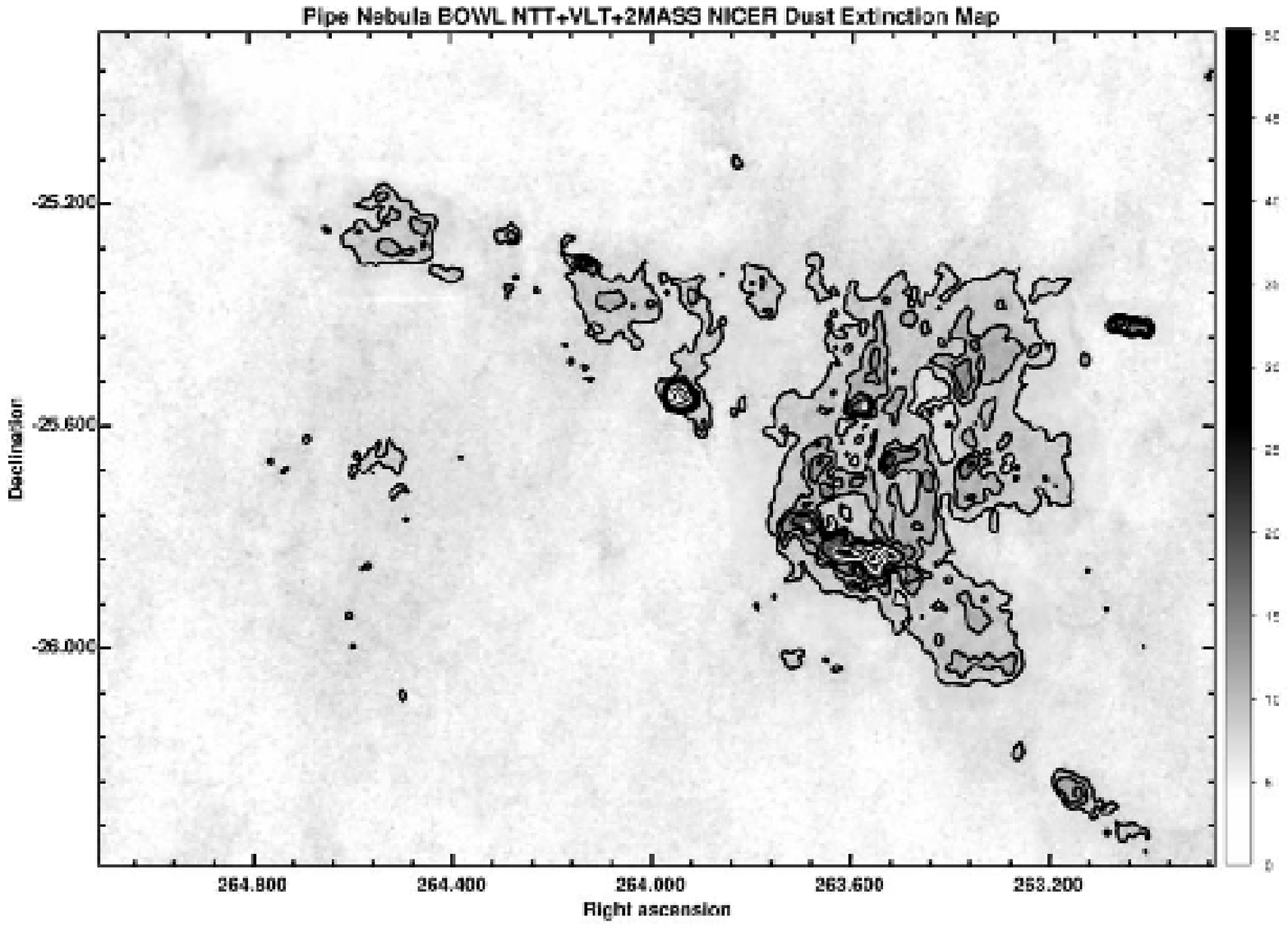}
\caption{Extinction map of the Pipe Bowl region at a spatial resolution of 20$\arcsec$. 
The white solid line contours mark levels of visual extinction at $A_V=7,10,12,15,20,25,
30,35,40\mathrm{\ and\ }50$ mag. \textit{Please see the electronic version of the Journal for a color version of this figure.} \label{fig:LMAP_BOWL}}
\end{figure*}

\begin{figure*}[ht]
\centering
\includegraphics[width=7.0in]{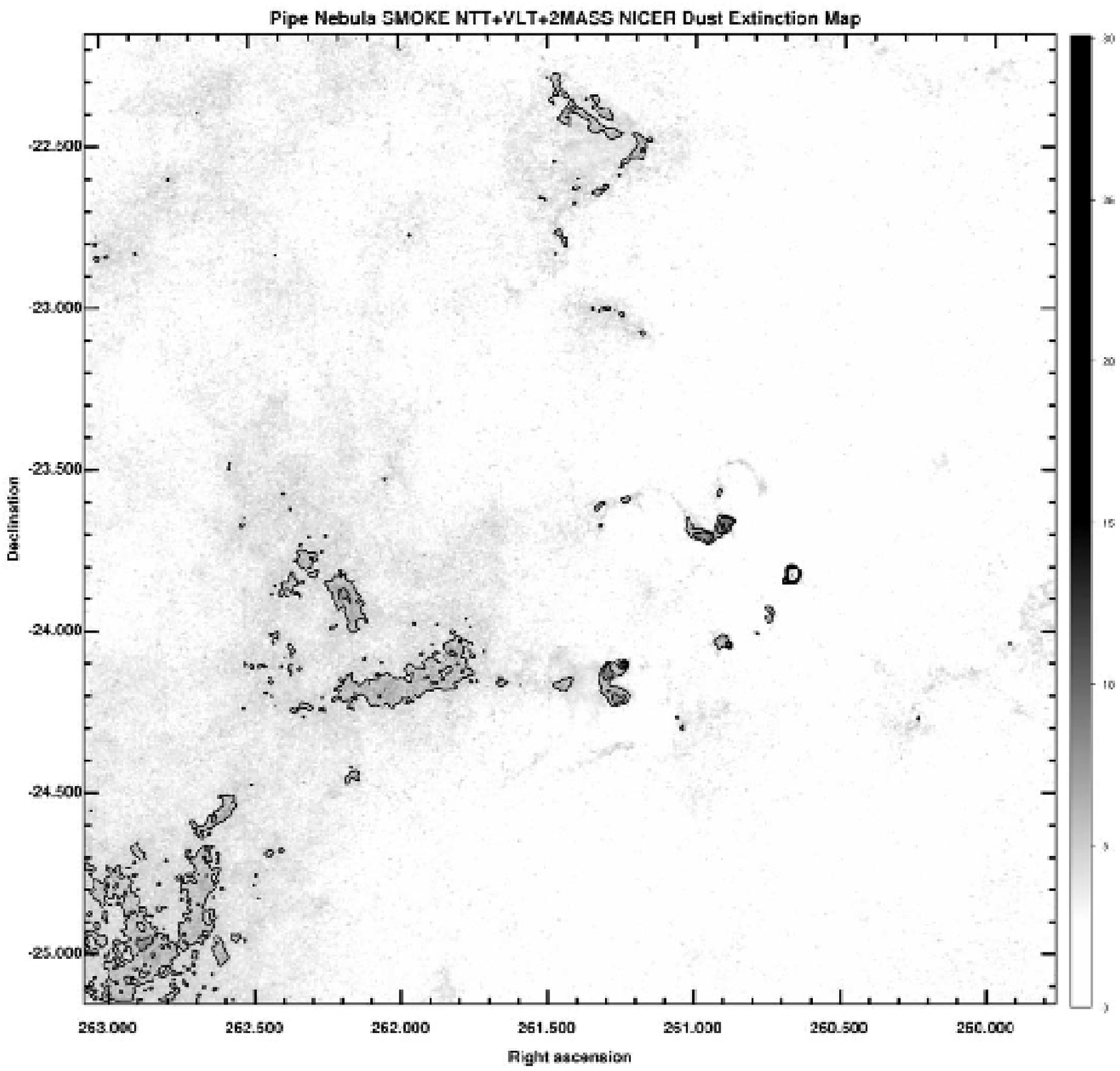}
\caption{Extinction map of the Pipe Smoke region at a spatial resolution of 20$\arcsec$. 
The white solid line contours mark levels of visual extinction at $A_V=3,5,7,10,15,20,
25\mathrm{\ and\ }30$ mag. \textit{Please see the electronic version of the Journal for a color version of this figure.} \label{fig:LMAP_SMOKE}}
\end{figure*}

\begin{figure}
\centering
\includegraphics[width=3.5in]{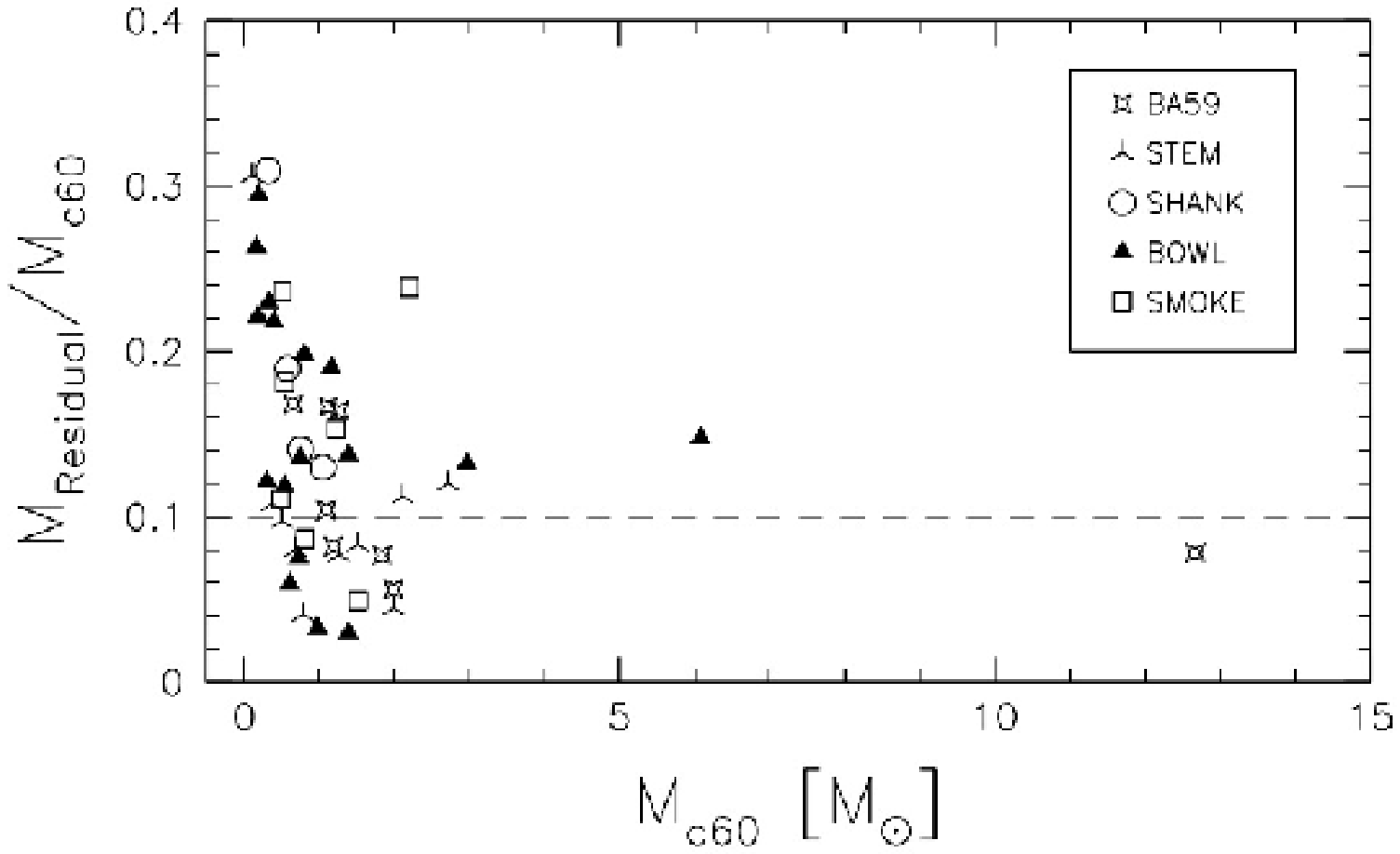}
\caption{Comparison of mass (M$_\odot$) for coincident peaks in the  residual
images and the corresponding low resolution (60$\arcsec$ Gaussian convolved) image. \label{fig:unsharp}}
\end{figure}

\begin{figure}
\centering
\includegraphics[width=3.5in]{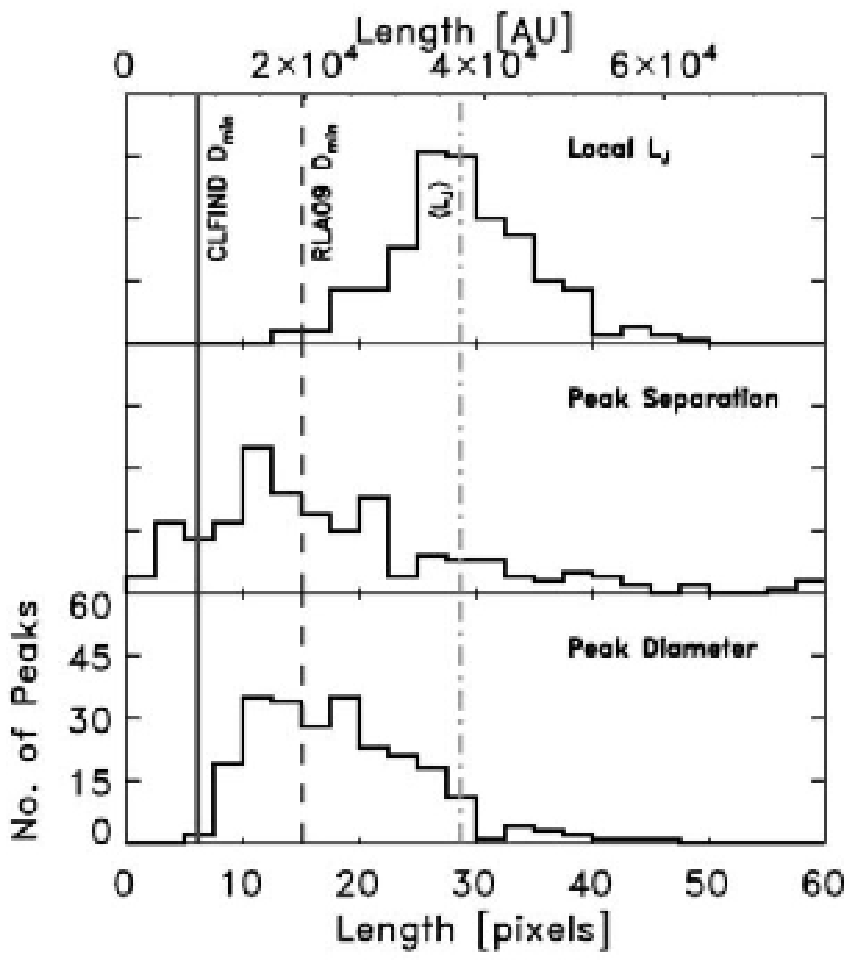}
\caption{The solid line histograms in each panel show, from bottom to top, the distributions of extinction peak diameters, extinction peaks nearest neighbor separations, and local Jeans length (derived from their average density and assuming $T=10$ K). The distribution are shown in a scale of pixel width units (one pixel width equals 10$\arcsec$). Also indicated at the top is the scale in astronomical units (AU). Vertical lines, solid, dashed and dot-dashed, from left to right, indicate the equivalent diameter for the minimum area detection threshold of CLF2D used in our analysis and in the RLA09 analysis, and the average Jeans length for all peaks, respectively. The range of values for core separations is as large as about $2.6\times10^5$ AU (200 pixels), but the range has been clipped to $7.8\times10^4$ AU (60 pixels) for clarity. \textit{Please see the electronic version of the Journal for a color version of this figure.} \label{fig:pixelcomp}}
\end{figure}

\begin{figure}
\centering
\includegraphics[width=3.5in]{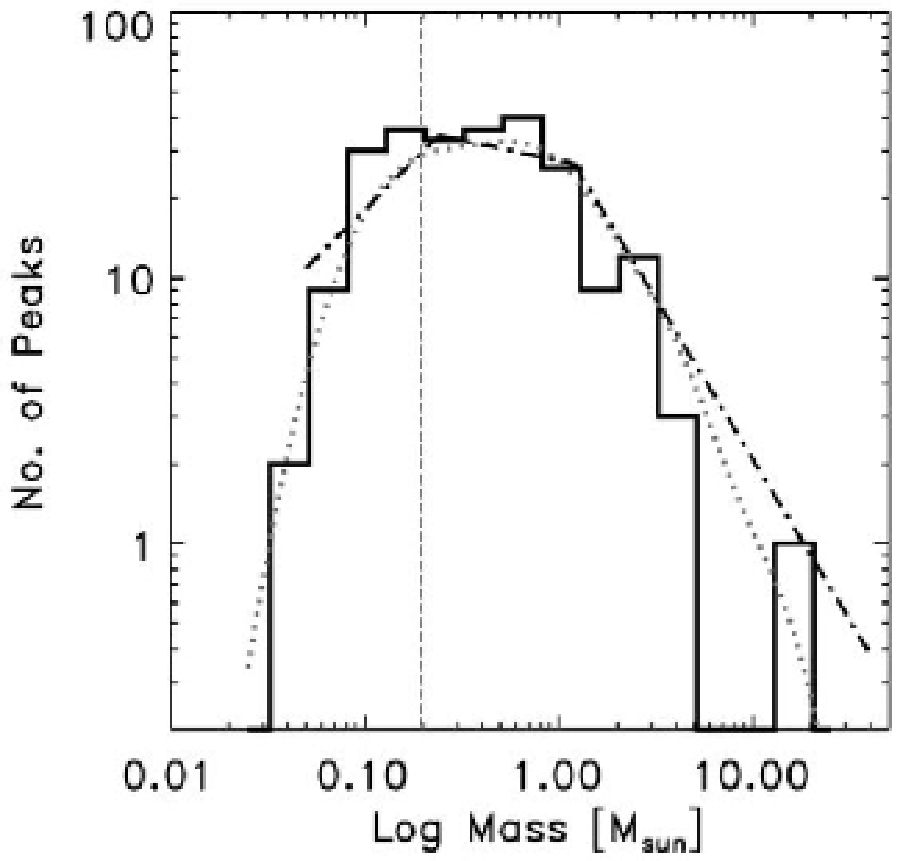}
\caption{ The histogram shows the distribution of mass values for all extinction peaks found in the large scale extinction maps. The light dotted line is the probability density function for the distribution, determined with a Gaussian kernel of width equal to $\log(M)=0.2$. The broken, dot-dashed line is the estimated IMF for the Trapezium cluster \citep{Muench:2002aa} scaled by a factor of 1.6 in mass and fit vertically to the PDF by a Chi-squared minimum fit. The thin, dotted line indicates the mass completeness limit. \textit{Please see the electronic version of the Journal for a color version of this figure.} \label{fig:peak_mass_distribution} 
}
\end{figure}

\begin{figure}
\centering
\includegraphics[width=3.5in]{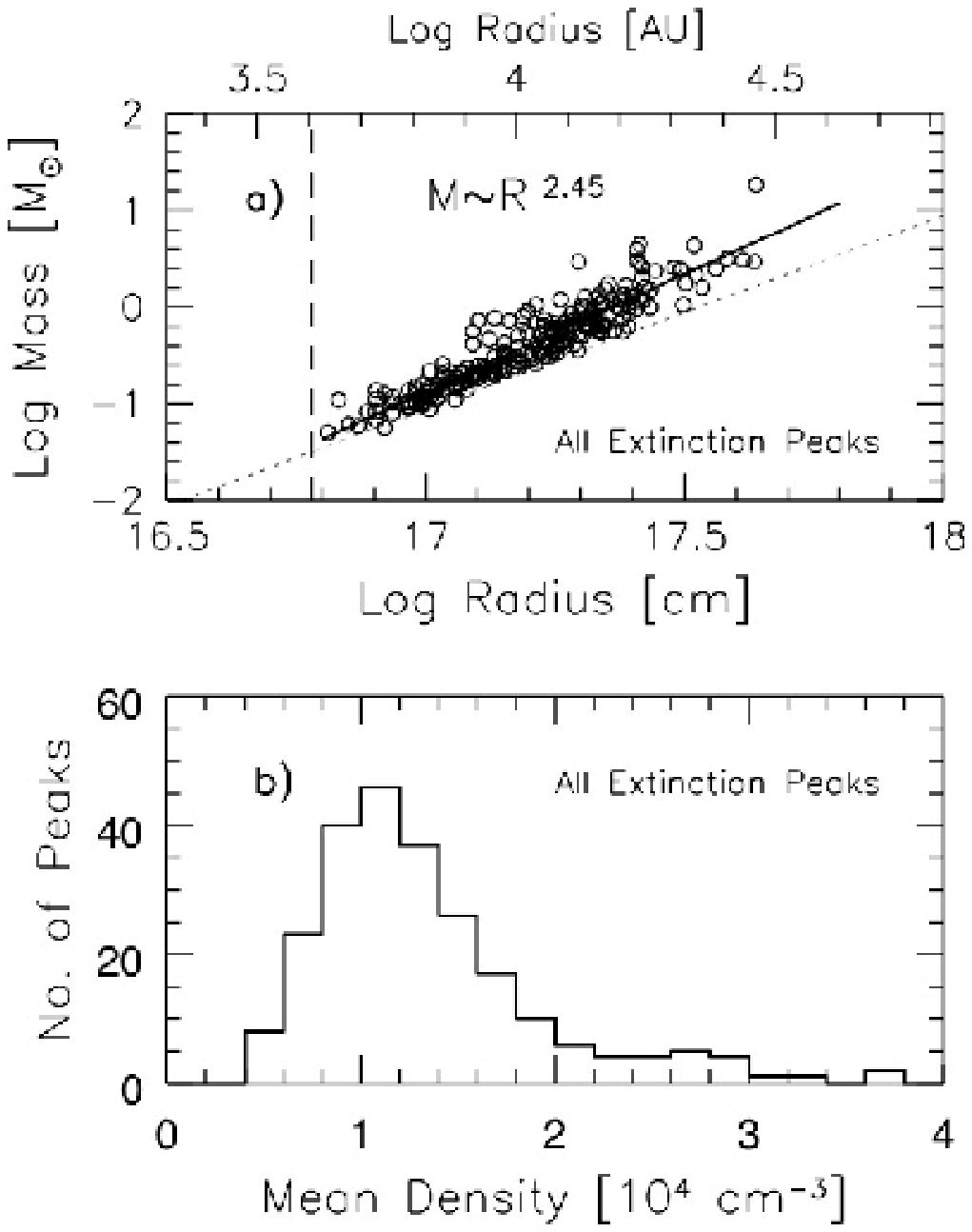}
\caption{{\it Top (a):} Mass-Radius relationship for all extinction peaks. The solid line is a least square fit to all data points. The dashed line marks the equivalent radius of the minimum detection area. The dotted line indicates the sensitivity threshold. {\it Bottom (b)}: Mean density distribution for all extinction peaks. \label{fig:mrplotdens}}
\end{figure}

\begin{figure}
\centering
\includegraphics[width=3.5in]{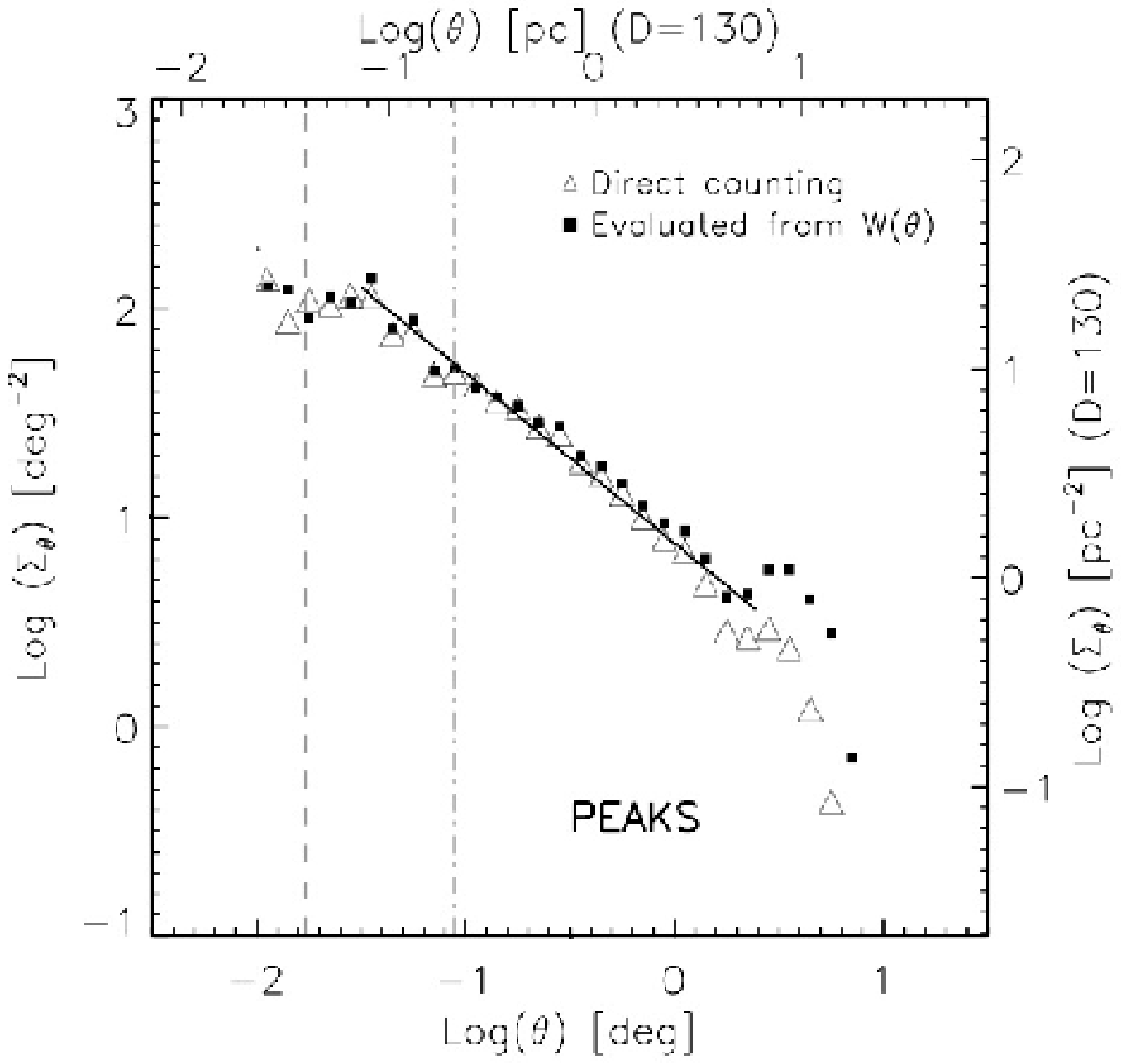}
\caption{The Mean Surface Density of Companions (MSDC), $\Sigma_\theta$, for all extinction peaks detected in the five large
scale maps. The triangle shaped data points represent the MSDC calculated from direct counting of feature separations in bins of equal width in the $\log{\theta}$ scale. The square filled points represent the MSDC recovered from the Two Point Correlation Function (TPCF), $W(\theta)$. The gray, vertical dot-dashed line is set at the median Local Jeans Length value. The gray, dotted line marks, for comparison, the minimum detection diameter for CLF2D. The solid line is a linear fit to the triangle shaped data points within the range $-1.5<\log{\theta}<0.4$. \label{fig:stt}}
\end{figure}

\begin{figure*}
\centering
\includegraphics[width=6.0in]{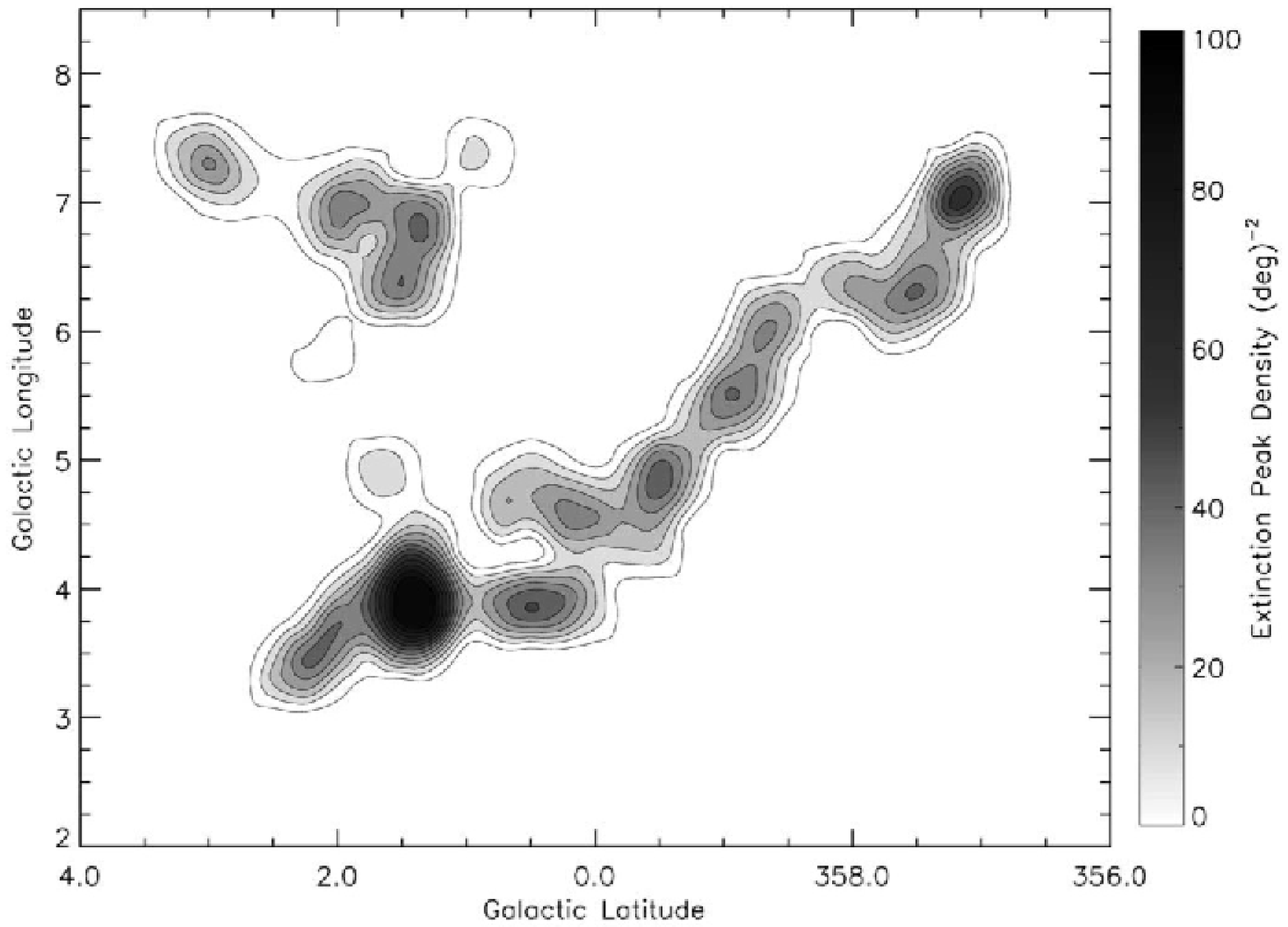}
\caption{A map of surface density of peaks across the Pipe Nebula. The map was constructed using a Nyquist sample square grid with a spatial resolution of 0.5$^\circ$. Contours levels are equally spaced by 10\% increments between the median and the maximum values of the peak surface density ($\Sigma _\theta$=2.0 and 100 $(^\circ)^{-2}$, respectively.) \textit{Please see the electronic version of the Journal for a color version of this figure.} \label{fig:csd1}}
\end{figure*}

\begin{figure}
\includegraphics[width=3.5in]{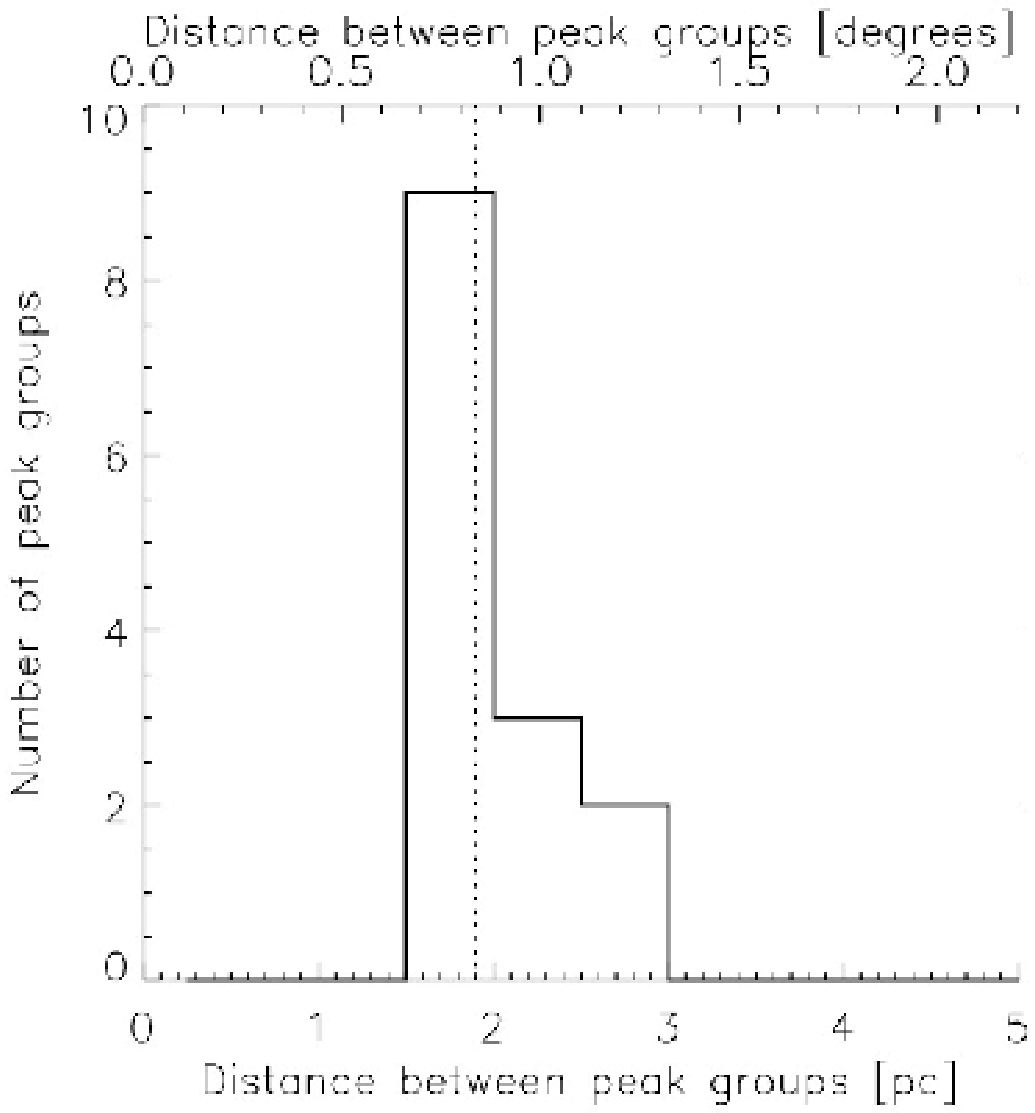}
\caption{Distribution of distances between local maxima of surface density of peaks (extinction peak groups). The dotted line shows the median value at 0.84$^\circ$ (1.9 pc). \label{fig:csd2}}
\end{figure}

\begin{figure}
\centering
\includegraphics[width=3.5in]{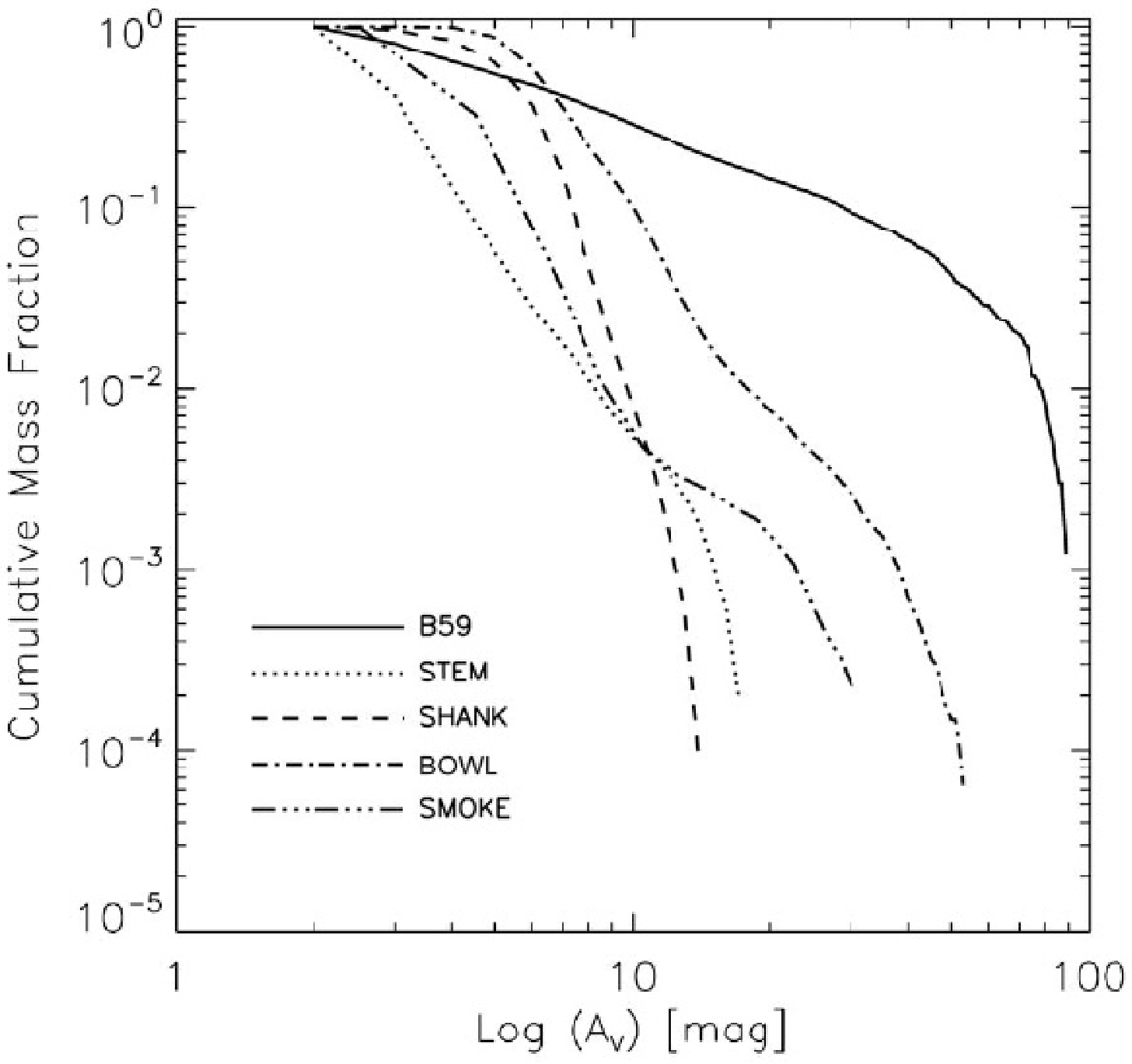}
\caption{Cumulative mass fraction as a function of extinction for the five large scale maps of the Pipe Nebula. \label{fig:cumul_mass}}
\end{figure}

We performed a series of Monte Carlo experiments, to determine the completeness of our peak mass sample. For each experiment we simulated a stellar field of $5\arcmin\times5\arcmin$ (equal to the SOFI FOV), with a brightness distribution and stellar surface density similar to those observed in the bulge field behind the Pipe ($8<K<21.0$ and 450 stars per sq. arcmin, respectively, as observed in the control field) and ran NICER over the field. Each star was then reddened according to a predetermined 2D column density distribution to simulate the peak. The mass of each peak, distributed accordingly over the column density profile, was pre-determined from the mass-size relationship. Then we added the resulting $A_V$ distribution at a random position on the background frame (original minus wavelet filtered) of the Shank region to form an embedded extinction feature. We processed the image with the wavelet filter and ran CLF2D. The completeness of the sample was then determined by counting how many peaks were recovered with a CLFD2D mass value falling within a 20\% tolerance range from the input value. We estimate that our method detects and measures correctly extinction peak masses for over $90\pm5$\% of peaks with masses above 0.2 M$_\odot$. below this limit, the detection rates drop to 80\% and then to 55\%, as shown in Figure \ref{fig:completeness}. This means that the last three bins in the histogram of Figure \ref{fig:peak_mass_distribution} would be off by a factor of less than 2. However, this also is a significant improvement over the detection rates in the map of Lombardi et al. for which the completeness limit is estimated to be 3 to 5 times larger.

\clearpage

\clearpage

\clearpage

\clearpage

\clearpage

\clearpage

\clearpage

\clearpage

\clearpage

\clearpage

\clearpage

\clearpage

\clearpage


\begin{deluxetable}{lcccccc}
\tablecolumns{5}
\tablewidth{0pc}
\tablecaption{Large Area Maps \label{tab:mapareas}} 
\tablehead{
\colhead{Map} &
\colhead{RA$_{min}$} &
\colhead{RA$_{max}$} &
\colhead{Dec$_{min}$} &
\colhead{Dec$_{max}$} \\
\cline{2-5} \\
\colhead{Region} &
\multicolumn{4}{c}{J2000} \\
} 
\startdata

Barnard 59 & 257.54996 & 258.30005 & -27.70047 & -27.25005\\
Stem       & 258.42673 & 261.24866 & -27.68037 & -26.42391\\
Shank      & 261.31604 & 263.11539 & -27.29518 & -25.69280\\
Bowl       & 262.87506 & 265.10638 & -26.39608 & -24.89208\\
Smoke      & 259.77020 & 263.08344 & -25.16239 & -22.15432\\

\enddata
\end{deluxetable}

\begin{deluxetable}{lcccc}
\tablecolumns{5}
\tablewidth{0pc}
\tablecaption{\texttt{CLUMPFIND-2D} Identifications in High-resolution Extinction Maps of the Pipe Nebula \label{tab:peaks}} 
\tablehead{
\colhead{Peak ID\tablenotemark{1}} &
\colhead{max $A_V$\tablenotemark{2}} & 
\colhead{$R_{eq}$} &
\colhead{Mass} &
\colhead{$\bar{n}$ } \\
\colhead{} &
\colhead{[mag]} &
\colhead{[pc]} &
\colhead{[M$_\odot$]} &
\colhead{[10$^4$ cm$^{-3}$]} \\
}
\startdata 

\cline{1-5}\\*
\multicolumn{5}{c}{Stem Region}\\*
\cline{1-5}\\*

R015a    &    6.4    &  0.054	 &  0.310  &  0.839 \\
R015b    &    6.3    &  0.032	 &  0.106  &  1.352 \\   
R016a    &   11.5    &  0.072	 &  1.328  &  1.558 \\   
R016b*   &   15.7    &  0.043	 &  0.789  &  4.314 \\   
R016c    &   12.9    &  0.036	 &  0.278  &  2.422 \\
R016d    &    8.0    &  0.035	 &  0.171  &  1.759 \\   
R016e    &    8.7    &  0.031	 &  0.150  &  2.076 \\   
R017     &    8.5    &  0.146	 &  3.194  &  0.453 \\   
R018a    &   10.5    &  0.089	 &  1.723  &  1.060 \\   
R018b*   &   10.9    &  0.084	 &  1.373  &  0.988 \\   
R020a    &    9.3    &  0.044	 &  0.244  &  1.288 \\   
R020b    &    8.5    &  0.037	 &  0.197  &  1.628 \\   
R020c    &    9.3    &  0.032	 &  0.149  &  1.875 \\   
R021a    &    8.5    &  0.087	 &  1.259  &  0.817 \\   
R021b    &    6.7    &  0.046	 &  0.230  &  1.031 \\   
R023     &    9.3    &  0.138	 &  3.311  &  0.544 \\
R027a    &    8.1    &  0.067	 &  0.737  &  1.092 \\   
R027b    &    7.6    &  0.073	 &  0.627  &  0.707 \\   
R027c    &    8.2    &  0.056	 &  0.540  &  1.339 \\   
R028     &    7.8    &  0.050	 &  0.259  &  0.903 \\   
R029a    &   13.1    &  0.129	 &  3.425  &  0.701 \\   
R029b    &   13.5    &  0.075	 &  1.669  &  1.759 \\   
R030     &   10.6    &  0.123	 &  2.676  &  0.637 \\   
R032     &    8.2    &  0.085	 &  1.215  &  0.865 \\   
R033a    &   12.8    &  0.094	 &  2.563  &  1.367 \\   
R033b    &    8.4    &  0.073	 &  0.865  &  0.981 \\   
R034a    &   24.6    &  0.111	 &  4.692  &  1.474 \\   
R034b    &   10.9    &  0.112	 &  2.327  &  0.705 \\   
R034c*   &   11.5    &  0.081	 &  1.456  &  1.186 \\   
R035a    &   21.8    &  0.088	 &  2.811  &  1.775 \\   
R035b*   &   13.4    &  0.073	 &  1.132  &  1.247 \\   
R036     &    8.6    &  0.066	 &  0.449  &  0.706 \\   
R037a    &    7.2    &  0.044	 &  0.203  &  1.022 \\   
R037b    &    6.9    &  0.041	 &  0.151  &  1.013 \\   
R038a    &    8.3    &  0.059	 &  0.375  &  0.771 \\   
R038b    &    6.5    &  0.039	 &  0.119  &  0.913 \\   
N001     &    8.7    &  0.055	 &  0.440  &  1.138 \\
N002     &    6.6    &  0.031	 &  0.097  &  1.364 \\   
N003     &    5.6    &  0.033	 &  0.106  &  1.266 \\   
N004a    &    7.6    &  0.040	 &  0.172  &  1.187 \\   
N004b    &    5.6    &  0.032	 &  0.114  &  1.423 \\   
N005     &    5.0    &  0.029	 &  0.087  &  1.558 \\   
N006     &    6.1    &  0.056	 &  0.366  &  0.905 \\   
N007a    &    6.6    &  0.035	 &  0.154  &  1.555 \\   
N007b    &    7.0    &  0.026	 &  0.090  &  2.138 \\   
N008     &    6.5    &  0.027	 &  0.073  &  1.561 \\   

\cline{1-5}\\*
\multicolumn{5}{c}{Shank Region}\\*
\cline{1-5}\\*

R039a    &    11.2   &  0.065  &  0.814 &  1.257 \\
R039b    &     9.7   &  0.035  &  0.147 &  1.481 \\
R040a    &    11.3   &  0.071  &  0.980 &  1.179 \\
R040b    &    11.0   &  0.059  &  0.742 &  1.565 \\
R040c    &     7.9   &  0.067  &  0.546 &  0.791 \\
R040d    &    10.1   &  0.057  &  0.426 &  1.004 \\
R040e    &     7.8   &  0.042  &  0.214 &  1.308 \\
R040f    &     7.4   &  0.036  &  0.149 &  1.366 \\
R041a    &     8.0   &  0.060  &  0.353 &  0.699 \\
R041b    &     7.4   &  0.058  &  0.324 &  0.696 \\
 R042    &     6.6   &  0.067  &  0.388 &  0.561 \\
 R043    &     8.2   &  0.074  &  0.580 &  0.638 \\
R045a    &    10.0   &  0.072  &  0.747 &  0.876 \\
R045b    &     9.1   &  0.075  &  0.687 &  0.720 \\
R046a    &    14.4   &  0.105  &  2.669 &  1.004 \\
R046b    &    10.8   &  0.087  &  1.388 &  0.899 \\
R046c    &    10.4   &  0.081  &  0.841 &  0.689 \\
 R047    &     9.6   &  0.056  &  0.371 &  0.907 \\
 R049    &     8.8   &  0.081  &  0.607 &  0.500 \\
R051a    &    11.3   &  0.079  &  1.089 &  0.952 \\
R051b    &    10.7   &  0.061  &  0.606 &  1.138 \\
R051c    &     9.9   &  0.057  &  0.334 &  0.787 \\
 R052    &    10.6   &  0.063  &  0.493 &  0.848 \\
R053a    &    15.7   &  0.045  &  0.502 &  2.519 \\
R053b    &    12.5   &  0.036  &  0.219 &  2.046 \\
R053c    &    11.4   &  0.046  &  0.333 &  1.552 \\
R053d    &    11.4   &  0.032  &  0.163 &  2.092 \\
R054a    &     8.5   &  0.044  &  0.265 &  1.423 \\
R054b    &     7.9   &  0.035  &  0.127 &  1.227 \\
R055a    &    10.2   &  0.070  &  0.696 &  0.881 \\ 
R055b    &    10.9   &  0.061  &  0.645 &  1.243 \\
R055c    &    10.2   &  0.048  &  0.358 &  1.434 \\
R057a    &    13.3   &  0.084  &  1.488 &  1.086 \\ 
R057b    &    11.0   &  0.080  &  1.035 &  0.867 \\ 
 R059    &     8.8   &  0.106  &  1.139 &  0.410 \\ 
N009a	 &     8.7   &  0.042  &  0.225 &  1.343 \\
N009b	 &     8.8   &  0.027  &  0.091 &  2.042 \\
N010	 &     8.6   &  0.078  &  0.608 &  0.554 \\
N011	 &     9.0   &  0.039  &  0.156 &  1.204 \\
N012a	 &     7.8   &  0.055  &  0.280 &  0.718 \\
N012b	 &     7.3   &  0.028  &  0.061 &  1.180 \\
N013	 &     8.9   &  0.033  &  0.114 &  1.397 \\
N014	 &    12.4   &  0.043  &  0.199 &  1.111 \\
N015	 &     8.4   &  0.034  &  0.095 &  0.995 \\
N016	 &    10.0   &  0.034  &  0.144 &  1.555 \\
N017	 &     7.8   &  0.035  &  0.119 &  1.197 \\
N018a	 &    10.4   &  0.047  &  0.250 &  1.073 \\
N018b	 &    10.2   &  0.046  &  0.217 &  1.020 \\
N018c	 &    10.0   &  0.042  &  0.202 &  1.176 \\
N018d	 &     9.7   &  0.033  &  0.105 &  1.247 \\

\cline{1-5}\\*
\multicolumn{5}{c}{Bowl Region}\\*
\cline{1-5}\\*

R060a   &  12.0   &  0.073  & 1.135  & 1.302 \\  
R060b   &   9.3   &  0.057  & 0.690  & 1.591 \\  
 R061   &   8.0   &  0.041  & 0.194  & 1.218 \\  
R064a   &  10.8   &  0.092  & 1.068  & 0.605 \\  
R064b   &  10.8   &  0.059  & 0.561  & 1.177 \\  
R064c*  &  10.7   &  0.058  & 0.459  & 0.999 \\  
 R070   &  12.4   &  0.070  & 0.636  & 0.814 \\  
R071a   &  47.4   &  0.087  & 4.812  & 3.172 \\  
R071b*  &  31.4   &  0.067  & 3.188  & 4.687 \\  
R071c   &  17.7   &  0.061  & 1.289  & 2.445 \\  
R071d   &  19.7   &  0.046  & 0.834  & 3.801 \\  
R071f*  &  13.4   &  0.061  & 0.597  & 1.114 \\  
R071g   &  12.2   &  0.041  & 0.252  & 1.661 \\  
R071h   &  10.9   &  0.034  & 0.159  & 1.717 \\  
 R072   &  11.9   &  0.071  & 0.900  & 1.131 \\  
R073a   &  15.0   &  0.065  & 0.927  & 1.473 \\  
R073b   &  12.3   &  0.072  & 0.857  & 1.018 \\  
R073c   &  10.9   &  0.053  & 0.375  & 1.103 \\  
R073d   &  10.1   &  0.043  & 0.208  & 1.150 \\  
R074a   &  15.6   &  0.042  & 0.601  & 3.501 \\  
R074b   &  13.4   &  0.055  & 0.730  & 1.933 \\  
R074c   &  14.5   &  0.042  & 0.448  & 2.728 \\  
R076a   &  18.7   &  0.055  & 1.141  & 2.916 \\  
R076b   &  18.7   &  0.053  & 0.982  & 2.838 \\  
R076c   &  11.5   &  0.051  & 0.411  & 1.322 \\  
R076d   &  11.5   &  0.049  & 0.344  & 1.268 \\  
R077a   &  14.9   &  0.052  & 0.726  & 2.285 \\  
R077b   &  13.8   &  0.065  & 0.704  & 1.142 \\  
R078a   &  11.9   &  0.060  & 0.482  & 0.955 \\  
R078b   &  11.6   &  0.058  & 0.471  & 1.020 \\  
R078c   &  10.1   &  0.060  & 0.449  & 0.893 \\  
R081a   &  13.4   &  0.080  & 1.281  & 1.085 \\  
R081b   &  11.5   &  0.053  & 0.379  & 1.129 \\  
R081c   &  11.6   &  0.039  & 0.200  & 1.527 \\  
R081d   &  11.1   &  0.032  & 0.133  & 1.762 \\  
R082a   &  14.6   &  0.067  & 0.958  & 1.408 \\  
R082b*  &  13.1   &  0.061  & 0.683  & 1.295 \\  
R082c   &  11.7   &  0.057  & 0.486  & 1.114 \\  
R082d   &  11.8   &  0.043  & 0.267  & 1.462 \\  
R083    &  12.4   &  0.047  & 0.281  & 1.194 \\  
R087a   &  20.7   &  0.073  & 1.431  & 1.600 \\  
R087b*  &  11.2   &  0.083  & 1.085  & 0.832 \\  
R088    &   8.3   &  0.060  & 0.363  & 0.739 \\  
R092    &  11.5   &  0.040  & 0.175  & 1.238 \\  
R094    &   8.7   &  0.061  & 0.566  & 1.044 \\  
R096    &  33.4   &  0.086  & 4.246  & 2.899 \\  
R101a   &   9.1   &  0.061  & 0.454  & 0.874 \\  
R101b   &  10.4   &  0.048  & 0.329  & 1.265 \\  
R103a   &  10.8   &  0.048  & 0.375  & 1.456 \\  
R103b   &   9.8   &  0.031  & 0.114  & 1.690 \\  
R108a   &  11.2   &  0.086  & 1.037  & 0.700 \\  
R108b   &  10.3   &  0.054  & 0.340  & 0.932 \\  
R110a   &  12.1   &  0.073  & 0.897  & 1.029 \\  
R110b   &   8.8   &  0.051  & 0.292  & 0.981 \\  
R114    &   9.5   &  0.074  & 0.717  & 0.790 \\  
R115a   &  10.7   &  0.057  & 0.476  & 1.085 \\  
R115b   &  11.9   &  0.037  & 0.233  & 1.973 \\  
R117    &  11.3   &  0.107  & 1.908  & 0.681 \\  
R118    &  10.4   &  0.085  & 1.386  & 0.960 \\  
R120a   &  10.7   &  0.051  & 0.542  & 1.744 \\ 
R120b   &   9.1   &  0.046  & 0.259  & 1.167 \\ 
R121    &  10.7   &  0.076  & 0.901  & 0.903 \\ 
R123    &   5.2   &  0.033  & 0.124  & 1.534 \\
N019    &   8.0   &  0.033  & 0.172  & 1.950 \\  
N020    &   8.4   &  0.024  & 0.066  & 2.186 \\  
N021    &   7.7   &  0.029  & 0.088  & 1.509 \\  

\cline{1-5}\\*
\multicolumn{5}{c}{Smoke Region}\\*
\cline{1-5}\\*

 R062   &   5.5  &  0.051  &  0.256   & 0.850 \\  
 R069   &   6.5  &  0.031  &  0.102   & 1.502 \\  
R075a   &   6.0  &  0.043  &  0.226   & 1.296 \\  
R075b   &   5.5  &  0.047  &  0.241   & 1.032 \\  
R075c   &   5.5  &  0.041  &  0.188   & 1.226 \\  
R079a   &   8.1  &  0.034  &  0.234   & 2.609 \\  
R079b   &   9.1  &  0.027  &  0.149   & 3.100 \\  
R079c   &   7.8  &  0.028  &  0.138   & 2.752 \\  
R079d   &   8.1  &  0.027  &  0.119   & 2.679 \\  
R079e   &   9.1  &  0.023  &  0.118   & 4.483 \\  
R080a   &   7.4  &  0.067  &  0.811   & 1.209 \\  
R080b   &  11.3  &  0.053  &  0.590   & 1.689 \\  
R084a   &   8.6  &  0.077  &  1.302   & 1.242 \\  
R084b   &   7.7  &  0.069  &  0.817   & 1.111 \\  
 R086   &  30.2  &  0.086  &  3.226   & 2.164 \\  
R090a   &   9.9  &  0.065  &  0.912   & 1.492 \\  
R090b   &   9.1  &  0.048  &  0.461   & 1.815 \\  
 R091   &  13.4  &  0.057  &  0.805   & 1.896 \\  
 R093   &   7.5  &  0.062  &  0.475   & 0.831 \\  
R095a   &   8.7  &  0.068  &  0.542   & 0.775 \\  
R095b   &   9.2  &  0.050  &  0.317   & 1.123 \\ 
 R097   &   6.9  &  0.059  &  0.388   & 0.815 \\  
R099a   &  11.1  &  0.082  &  1.811   & 1.434 \\  
R099b   &   7.1  &  0.088  &  1.230   & 0.765 \\  
R100a   &  10.7  &  0.088  &  2.158   & 1.339 \\  
R100b   &   7.2  &  0.061  &  0.553   & 1.028 \\  
 R102   &   6.2  &  0.072  &  0.690   & 0.815 \\  
 R104   &   9.7  &  0.083  &  0.914   & 0.690 \\  
R105a   &   6.5  &  0.063  &  0.488   & 0.840 \\  
R105b   &   6.3  &  0.035  &  0.170   & 1.745 \\  
 R106   &   8.9  &  0.056  &  0.355   & 0.879 \\  
R111a   &   8.0  &  0.061  &  0.581   & 1.072 \\  
R111b   &   5.5  &  0.040  &  0.220   & 1.538 \\  
R111c   &   6.1  &  0.029  &  0.110   & 1.936 \\  
R112a   &   7.5  &  0.062  &  0.528   & 0.937 \\  
R112b   &   6.3  &  0.039  &  0.182   & 1.388 \\  
R116    &   9.5  &  0.115  &  1.728   & 0.492 \\  
R122    &   7.2  &  0.062  &  0.428   & 0.754 \\  
R124    &   8.1  &  0.083  &  0.674   & 0.513 \\  
R127a   &   6.2  &  0.045  &  0.226   & 1.072 \\  
R127b   &   6.9  &  0.044  &  0.222   & 1.133 \\  
R128a   &   7.7  &  0.036  &  0.191   & 1.694 \\   
R128b   &   6.5  &  0.033  &  0.132   & 1.522 \\   
R129    &   6.7  &  0.039  &  0.162   & 1.241 \\   
R130a   &   7.2  &  0.046  &  0.262   & 1.161 \\   
R130b   &   7.2  &  0.043  &  0.204   & 1.154 \\   
R131a   &   8.6  &  0.048  &  0.254   & 1.002 \\   
R131b   &   8.9  &  0.040  &  0.229   & 1.572 \\   
N022a   &   6.0  &  0.039  &  0.179   & 1.330 \\  
N022b   &   7.9  &  0.030  &  0.108   & 1.705 \\  
 N023   &   4.7  &  0.073  &  0.596   & 0.677 \\  
 N024   &   6.2  &  0.048  &  0.234   & 0.956 \\  
 N025   &   5.1  &  0.031  &  0.101   & 1.407 \\  
 N026   &  10.2  &  0.029  &  0.127   & 2.234 \\  
 N027   &   8.6  &  0.022  &  0.054   & 2.310 \\  
 N028   &   7.0  &  0.052  &  0.291   & 0.917 \\  
 N029   &   5.2  &  0.034  &  0.124   & 1.367 \\  
 N030   &   4.8  &  0.025  &  0.064   & 1.840 \\  

\enddata
\tablenotetext{1}{Peak identifications follow the numbering used in the 
list of RLA09. Peaks located by less than one Jeans length from a coincident RLA09 position are listed with the core number followed by
letters (a,b,c, etc), which list peaks alphabetically from the largest
to the smallest mass; an asterisk indicates when a peak in a group is also
separated by less than 0.12 km/s in radial velocity with respect to the
main position; peaks identified with `N' were previously undetected.}
\tablenotetext{2}{Peak extinction values in this list are those
listed by CLF2D for the wavelet-filtered maps (which have a background value of zero).}
\end{deluxetable}

\begin{deluxetable}{lccccccc}
\tablecolumns{3}
\tablewidth{0pc}
\tablecaption{Cores from RLA09 not detected or not included in Large Scale Maps\label{tab:missing}} 
\tablehead{
\colhead{Region} &
\colhead{Undetected\tablenotemark{a}} &
\colhead{Not included\tablenotemark{b}} &
}
\startdata
B59    & 14			  	& 01,10,12 \\
Stem   & 22, 25, 26, 31		  	& --\\
Shank  & 44, 50, 56		  	& 24, 66\\
Bowl   & 65, 67, 85, 89, 113, 125, 126	& --\\
Smoke  & 62 ,63, 68, 98, 116, 132 	& 48, 58, 133, 134\\
\enddata
\tablenotetext{a}{These objects broke down and were not detected by CLUMPFIND-2D}
\tablenotetext{b}{These objects fall outside the regions covered in the maps}
\end{deluxetable}


\begin{thebibliography}{43}
\expandafter\ifx\csname natexlab\endcsname\relax\def\natexlab#1{#1}\fi

\bibitem[{{Alves} {et~al.}(2007){Alves}, {Lombardi}, \& {Lada}}]{Alves:2007aa}
{Alves}, J., {Lombardi}, M., \& {Lada}, C.~J. 2007, \aap, 462, L17

\bibitem[{{Alves} {et~al.}(2001){Alves}, {Lada}, \& {Lada}}]{Alves:2001aa}
{Alves}, J.~F., {Lada}, C.~J., \& {Lada}, E.~A. 2001, \nat, 409, 159

\bibitem[{{Bergin} \& {Tafalla}(2007)}]{Bergin:2007aa}
{Bergin}, E.~A. \& {Tafalla}, M. 2007, \araa, 45, 339

\bibitem[{{Bertin} \& {Arnouts}(1996)}]{Bertin:1996aa}
{Bertin}, E. \& {Arnouts}, S. 1996, \aaps, 117, 393

\bibitem[{{Bijaoui} {et~al.}(1997){Bijaoui}, {Ru{\'e}}, \&
  {Vandame}}]{Bijaoui:1997aa}
{Bijaoui}, A., {Ru{\'e}}, F., \& {Vandame}, B. 1997, in Data Analysis in
  Astronomy IV, ed. V.~{Di Gesu}, M.~J.~B. {Duff}, A.~{Heck}, M.~C.
  {Maccarone}, L.~{Scarsi}, \& H.~U. {Zimmerman}, 337

\bibitem[{{Brooke} {et~al.}(2007)}]{Brooke:2007aa}
{Brooke}, T.~Y., {Huard}, T.~L., {Bourke}, et~al. 2007, \apj, 655, 364

\bibitem[{{Fabian} {et~al.}(2006){Fabian}, {Sanders}, {Taylor}, {Allen},
  {Crawford}, {Johnstone}, \& {Iwasawa}}]{Fabian:2006aa}
{Fabian}, A.~C., {Sanders}, J.~S., {Taylor}, G.~B., {Allen}, S.~W., {Crawford},
  C.~S., {Johnstone}, R.~M., \& {Iwasawa}, K. 2006, \mnras, 366, 417

\bibitem[Franco et al.(2010)]{Franco:2010aa} Franco, G.~A.~P., Alves, 
F.~O., \& Girart, J.~M.\ 2010, \apj, 723, 146 

\bibitem[{{Forbrich} {et~al.}(2009){Forbrich}, {Lada}, {Muench}, {Alves}, \&
{Lombardi}}]{Forbrich:2009ab} 
{Forbrich}, J., {Lada}, C.~J., {Muench}, A.~A., {Alves}, J., \& {Lombardi}, M.
  2009, \apj, 704, 292

\bibitem[{{G\'omez} {et~al.}(1993){G\'omez}, {Hartmann}, {Kenyon}, \&
  {Hewett}}]{Gomez:1993aa}
{G\'omez}, M., {Hartmann}, L., {Kenyon}, S.~J., \& {Hewett}, R. 1993, \aj, 105,
  1927

\bibitem[{{Goodman} {et~al.}(1998){Goodman}, {Barranco}, {Wilner}, \&
  {Heyer}}]{Goodman:1998aa}
{Goodman}, A.~A., {Barranco}, J.~A., {Wilner}, D.~J., \& {Heyer}, M.~H. 1998,
  Astrophysical Letters Communications, 37, 109

\bibitem[{{Hartmann}(2002)}]{Hartmann:2002bj}
{Hartmann}, L. 2002, \apj, 578, 914

\bibitem[{{Hewett}(1982)}]{Hewett:1982aa}
{Hewett}, P.~C. 1982, \mnras, 201, 867

\bibitem[{{Kainulainen} {et~al.}(2009){Kainulainen}, {Lada}, {Rathborne}, \&
  {Alves}}]{Kainulainen:2009aa}
{Kainulainen}, J., {Lada}, C.~J., {Rathborne}, J.~M., \& {Alves}, J.~F. 2009,
  \aap, 497, 399

\bibitem[{{Kandori} {et~al.}(2005){Kandori}, {Nakajima}, {Tamura}, {Tatematsu},
  {Aikawa}, {Naoi}, {Sugitani}, {Nakaya}, {Nagayama}, {Nagata}, {Kurita},
  {Kato}, {Nagashima}, \& {Sato}}]{Kandori:2005aa}
{Kandori}, R., {Nakajima}, Y., {Tamura}, M., {Tatematsu}, K., {Aikawa}, Y.,
  {Naoi}, T., {Sugitani}, K., {Nakaya}, H., {Nagayama}, T., {Nagata}, T.,
  {Kurita}, M., {Kato}, D., {Nagashima}, C., \& {Sato}, S. 2005, \aj, 130, 2166


\bibitem[{{Kraus} \& {Hillenbrand}(2008)}]{Kraus:2008aa}
{Kraus}, A.~L. \& {Hillenbrand}, L.~A. 2008, \apjl, 686, L111

\bibitem[{{Lada} {et~al.}(2004){Lada}, {Huard}, {Crews}, \&
  {Alves}}]{Lada:2004aa}
{Lada}, C.~J., {Huard}, T.~L., {Crews}, L.~J., \& {Alves}, J.~F. 2004, \apj,
  610, 303

\bibitem[{{Lada} {et~al.}(1994){Lada}, {Lada}, {Clemens}, \&
  {Bally}}]{Lada:1994aa}
{Lada}, C.~J., {Lada}, E.~A., {Clemens}, D.~P., \& {Bally}, J. 1994, \apj, 429,
  694

\bibitem[{{Lada} {et~al.}(2008){Lada}, {Muench}, {Rathborne}, {Alves}, \&
  {Lombardi}}]{Lada:2008aa}
{Lada}, C.~J., {Muench}, A.~A., {Rathborne}, J., {Alves}, J.~F., \& {Lombardi},
  M. 2008, \apj, 672, 410

\bibitem[{{Lada} {et~al.}(2009){Lada}, {Lombardi}, \& {Alves}}]{Lada:2009ab}
{Lada}, C.~J., {Lombardi}, M., \& {Alves}, J.~F. 2009, \apj, 703, 52

\bibitem[{{Larson}(1995)}]{Larson:1995sr}
{Larson}, R.~B. 1995, \mnras, 272, 213

\bibitem[{{Levine}(2006)}]{Levine:2006ab}
{Levine}, J. 2006, PhD thesis, University of Florida

\bibitem[{{Lombardi} \& {Alves}(2001)}]{Lombardi:2001aa}
{Lombardi}, M. \& {Alves}, J. 2001, \aap, 377, 1023

\bibitem[{{Lombardi} {et~al.}(2006){Lombardi}, {Alves}, \&
  {Lada}}]{Lombardi:2006aa}
{Lombardi}, M., {Alves}, J., \& {Lada}, C.~J. 2006, \aap, 454, 781

\bibitem[{{Lombardi} {et~al.}(2008){Lombardi}, {Lada}, \&
  {Alves}}]{Lombardi:2008aa}
{Lombardi}, M., {Lada}, C.~J., \& {Alves}, J. 2008, \aap, 489, 143

\bibitem[{{Moriarty-Schieven} {et~al.}(2006){Moriarty-Schieven}, {Johnstone},
  {Bally}, \& {Jenness}}]{Moriarty-Schieven:2006aa}
{Moriarty-Schieven}, G.~H., {Johnstone}, D., {Bally}, J., \& {Jenness}, T.
  2006, \apj, 645, 357

\bibitem[{{Muench} {et~al.}(2002){Muench}, {Lada}, {Lada}, \&
  {Alves}}]{Muench:2002aa}
{Muench}, A.~A., {Lada}, E.~A., {Lada}, C.~J., \& {Alves}, J. 2002, \apj, 573,
  366

\bibitem[{{Muench} {et~al.}(2007){Muench}, {Lada}, {Rathborne}, {Alves}, \&
  {Lombardi}}]{Muench:2007aa}
{Muench}, A.~A., {Lada}, C.~J., {Rathborne}, J.~M., {Alves}, J.~F., \&
  {Lombardi}, M. 2007, \apj, 671, 1820

\bibitem[{{Onishi} {et~al.}(1999){Onishi}, {Kawamura}, {Abe}, {Yamaguchi},
  {Saito}, {Moriguchi}, {Mizuno}, {Ogawa}, \& {Fukui}}]{Onishi:1999aa}
{Onishi}, T., {Kawamura}, A., {Abe}, R., {Yamaguchi}, N., {Saito}, H.,
  {Moriguchi}, Y., {Mizuno}, A., {Ogawa}, H., \& {Fukui}, Y. 1999, \pasj, 51,
  871

\bibitem[{{Peebles}(1973)}]{Peebles:1973aa}
{Peebles}, P.~J.~E. 1973, \apj, 185, 413

\bibitem[{{Rathborne} {et~al.}(2009{\natexlab{a}}){Rathborne}, {Lada},
  {Muench}, {Alves}, {Kainulainen}, \& {Lombardi}}]{Rathborne:2009aa}
{Rathborne}, J.~M., {Lada}, C.~J., {Muench}, A.~A., {Alves}, J.~F.,
  {Kainulainen}, J., \& {Lombardi}, M. 2009{\natexlab{a}}, \apj, 699, 742

\bibitem[{{Rathborne} {et~al.}(2008){Rathborne}, {Lada}, {Muench}, {Alves}, \&
  {Lombardi}}]{Rathborne:2008aa}
{Rathborne}, J.~M., {Lada}, C.~J., {Muench}, A.~A., {Alves}, J.~F., \&
  {Lombardi}, M. 2008, \apjs, 174, 396

\bibitem[{{Rathborne} {et~al.}(2009{\natexlab{b}}){Rathborne}, {Lada}, {Walsh},
  {Saul}, \& {Butner}}]{Rathborne:2009ab}
{Rathborne}, J.~M., {Lada}, C.~J., {Walsh}, W., {Saul}, M., \& {Butner}, H.~M.
  2009{\natexlab{b}}, \apj, 690, 1659

\bibitem[{{Rom\'an-Z\'u\~niga}(2006)}]{Roman-Zuniga:2006aa}
{Rom\'an-Z\'u\~niga}, C.~G. 2006, PhD thesis, University of Florida

\bibitem[{{Rom{\'a}n-Z{\'u}{\~n}iga} {et~al.}(2009){Rom{\'a}n-Z{\'u}{\~n}iga},
  {Lada}, \& {Alves}}]{Roman-Zuniga:2009aa}
{Rom{\'a}n-Z{\'u}{\~n}iga}, C.~G., {Lada}, C.~J., \& {Alves}, J.~F. 2009, \apj,
  704, 183

\bibitem[{{Rom{\'a}n-Z{\'u}{\~n}iga} {et~al.}(2007){Rom{\'a}n-Z{\'u}{\~n}iga},
  {Lada}, {Muench}, \& {Alves}}]{Roman-Zuniga:2007aa}
{Rom{\'a}n-Z{\'u}{\~n}iga}, C.~G., {Lada}, C.~J., {Muench}, A., \& {Alves},
  J.~F. 2007, \apj, 664, 357

\bibitem[{{Ru{\'e}} \& {Bijaoui}(1997)}]{Rue:1997aa}
{Ru{\'e}}, F. \& {Bijaoui}, A. 1997, Experimental Astronomy, 7, 129

\bibitem[{{Schnee} {et~al.}(2010){Schnee}, {Enoch}, {Johnstone}, {Culverhouse},
  {Leitch}, {Marrone}, \& {Sargent}}]{Schnee:2010aa}
{Schnee}, S., {Enoch}, M., {Johnstone}, D., {Culverhouse}, T., {Leitch}, E.,
  {Marrone}, D.~P., \& {Sargent}, A. 2010, \apj, 718, 306

\bibitem[{{Simon}(1997)}]{Simon:1997aa}
{Simon}, M. 1997, \apjl, 482, L81+

\bibitem[{{Tinney} {et~al.}(2003){Tinney}, {Burgasser}, \&
  {Kirkpatrick}}]{Tinney:2003aa}
{Tinney}, C.~G., {Burgasser}, A.~J., \& {Kirkpatrick}, J.~D. 2003, \aj, 126,
  975

\bibitem[{{Williams} {et~al.}(1994){Williams}, {de Geus}, \&
  {Blitz}}]{Williams:1994aa}
{Williams}, J.~P., {de Geus}, E.~J., \& {Blitz}, L. 1994, \apj, 428, 693

\end{thebibliography}
\end{document}